\documentclass[article]{jss}

\usepackage{thumbpdf,lmodern}

\usepackage{framed}
\usepackage{bm,amssymb,amsmath,amsthm}
\usepackage{epstopdf}
\usepackage{verbatim}
\usepackage{multirow}
\usepackage{bbm}

\newcommand{\class}[1]{`\code{#1}'}
\newcommand{\fct}[1]{\code{#1()}}

\author{Xueying Tang\\University of Arizona
   \And Susu Zhang\\Columbia University
   \And Zhi Wang\\Columbia University
   \AND Jingchen Liu\\Columbia University
   \And Zhiliang Ying\\Columbia University}
\Plainauthor{Xueying Tang, Susu Zhang, Zhi Wang, Jingchen Liu, Zhiliang Ying}

\title{\pkg{ProcData}: An \proglang{R} Package for Process Data Analysis}
\Plaintitle{ProcData: An R Package for Process Data Analysis}
\Shorttitle{Process Data Analysis in \proglang{R}}

\Abstract{
Process data refer to data recorded in the log files of computer-based items. These data, represented as timestamped action sequences, keep track of respondents' response processes of solving the items. Process data analysis aims at enhancing educational assessment accuracy and serving other assessment purposes by utilizing the rich information contained in response processes.
The \proglang{R} package \pkg{ProcData} presented in this article is designed to provide tools for processing, describing, and analyzing process data. We define an \proglang{S}3 class \class{proc} for organizing process data and extend generic methods \code{summary} and \code{print} for \class{proc}. 
Two feature extraction methods for process data are implemented in the package for compressing information in the irregular response processes into regular numeric vectors. \pkg{ProcData} also provides functions for fitting and making predictions from a neural-network-based sequence model. These functions call relevant functions in package \pkg{keras} for constructing and training neural networks. In addition, several response process generators and a real dataset of response processes of the climate control item in the 2012 Programme for International Student Assessment are included in the package.
}

\Plainkeywords{process data analysis, multidimensional scaling, autoencoder, sequence model}

\Address{
  Xueying Tang\\
  Deparment of Mathematics\\
  University of Arizona\\
  Address: 617 N. Santa Rita Avenue, Tucson, AZ 85721\\
  E-mail: \email{xytang@math.arizona.edu}\\
  URL: \url{http://websites.math.arizona.edu/xueyingtang/}\\
  \\
  \\
  Susu Zhang, Zhi Wang, Jingchen Liu, Zhiliang Ying\\
  Department of Statistics\\
  Columbia University\\
  Address: 1255 Amsterdam Avenue, New York, NY 10027\\
}

\begin{document}



\section[Introduction]{Introduction} \label{sec:intro}

With the advancement of technology, computer-based assessments have become popular in measuring complex human skills such as problem solving skills. In these assessments, participants are often asked to fulfill one or more real life tasks in a simulated environment. As a participant interacts with a computer to complete the tasks, the entire interaction process will be recorded in log files. Each recorded process includes the actions such as mouse clicks and keystrokes taken by the participant and the timestamps at which these actions took place. Such data describe the process of responding to an item and are thus called response process data, or in short, process data.

\begin{figure}[htb]
\includegraphics[width=\textwidth]{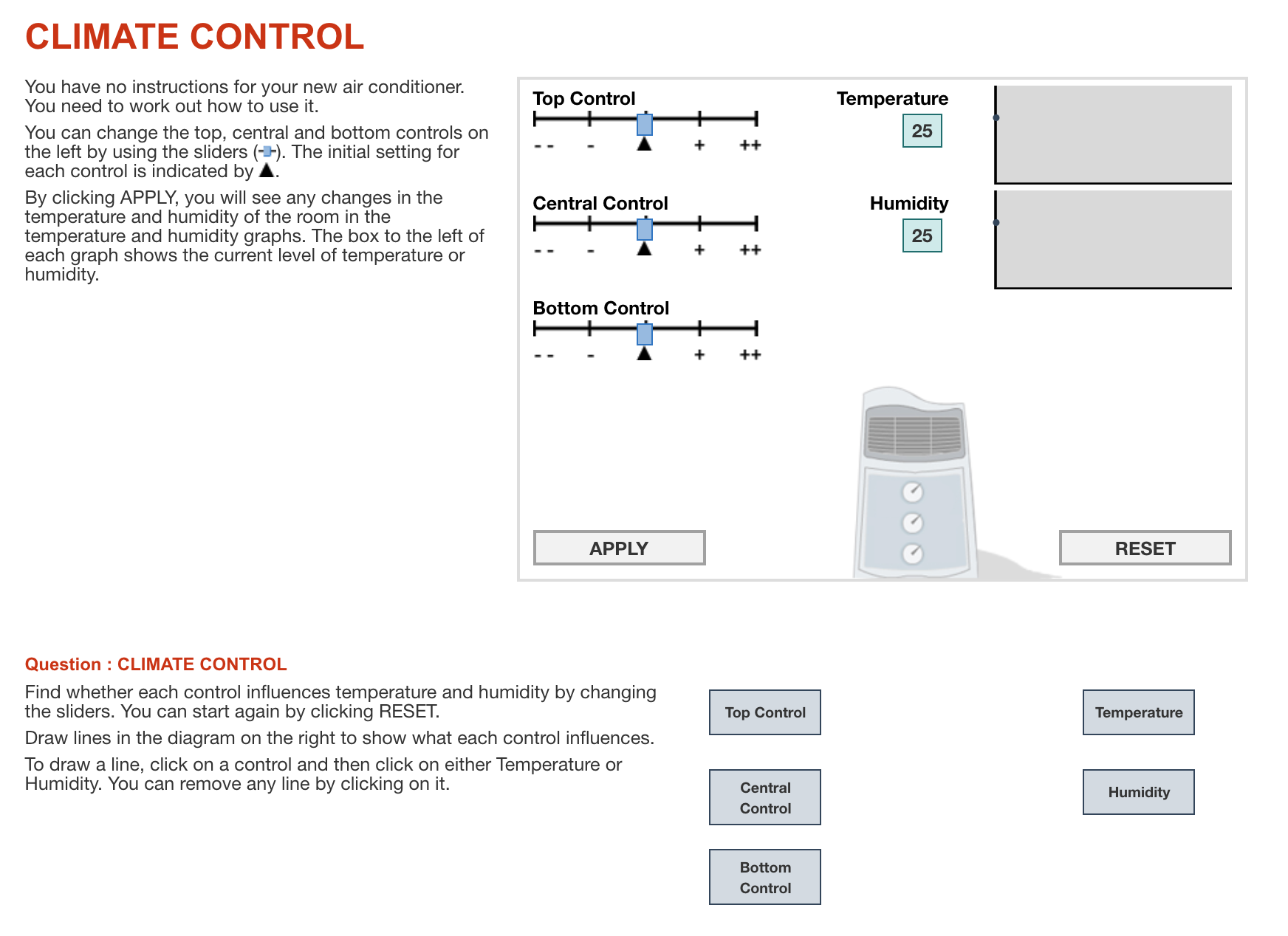}
\caption{Climate control item in PISA 2012.}\label{fig:cc_interface}
\end{figure}

The climate control item described below is an example of computer-based items from which response process data are collected. 
It belongs to the 2012 survey in the Programme for International Student Assessment (PISA) for assessing students' problem solving skills. Figure \ref{fig:cc_interface} is a screenshot of the item interface. In the simulated environment, there is a new air conditioner with no instructions.
Students are asked to figure out which climate variable, temperature or humidity, that each of the three controls on the air conditioner influences. They can slide the control bars through the simulation interface and read how the temperature and humidity change.
In the exploration process, how the controls are moved and buttons are clicked is recorded in the log files. For example, if a student clicked the ``APPLY'' button after moving the top control to ``+'' and the middle control to ``$--$'', then action ``1\_-2\_0'' along with the time elapsed since the start of the item, say 5.4 seconds, are recorded. A recorded process with actions ``1\_0\_0'', ``RESET'', ``0\_0\_-2'' and timestamps 4.9, 6.3, 10.6 indicates that the student moved the top control to ``+'' and clicked ``APPLY'' 4.9 seconds after the item started. The positions of the three controls were reset by a click of the ``RESET'' button 1.4 seconds later. Then the bottom control was moved to ``$--$'' and ``APPLY'' was clicked again 10.6 seconds after the item started. This sequence of actions constitutes a response process of the climate control item.

Based on the above description, each observation $O = (S, T)$ of process data consists of two sequences, an action sequence $S = (s_1, \ldots, s_L)$ and a timestamp sequence $T = (t_1, \ldots, t_L)$. The action sequence $S$ records the actions taken by the respondent to solve the item in order. Each element is an action in the set of possible actions $\mathcal{A}=\{a_1, \ldots, a_N\}$ for the item. Elements in the timestamp sequence $T$ record the time of actions in $S$ from inception and therefore $0 \leq t_1 \leq t_2 \leq \cdots \leq t_L$. 
For a set of response processes $O_1, \ldots, O_n$ of $n$ respondents, the length of a process is likely to vary across observations. We write $O_i = (S_i, T_i)$, $S_i = (s_{i1}, \ldots, s_{iL_i})$, and $T_i = (t_{i1}, \ldots, t_{iL_i})$, where $L_i$ denotes the length of the response process for respondent $i$, $i=1, \ldots, n$. 

One of the goals of process data analysis is to study how students' response processes are related to their observed characteristics such as demographics and education history and to their latent traits such as problem solving ability and personality. Traditionally, the relation between covariates and response variables is explored through linear or generalized linear models. The effect of latent attributes on item responses, such as final polytomous or dichotomous scores, can be modeled through latent variable models, including item response theory models \citep{lord1968statistical, lord1980applications} or cognitive diagnosis models \citep{rupp2010diagnostic}. However, because of the complex structure of response processes, most of the traditional methods cannot be directly applied to process data.


In the past few years, various methods have been proposed for analyzing process data. \citet{he2016analyzing} and \citet{qiao2019data} found that n-grams of action sequences are useful for predicting the item performance and clustering respondents. \citet{chen2019statistical} analyzed process data through an event history analysis approach. More recently, two feature extraction methods \citep{tang2019mds, tang2019seq2seq} are developed to automatically construct informative features from process data. One is based on multidimensional scaling while the other is based on a type of neural networks called autoencoder. Besides autoencoders, other neural networks such as recurrent neural networks have also been found potentially useful for analyzing process data \citep{tang2016deep}. The detailed response information contained in process data has been utilized to enhance education measurement, compare behavior patterns in successful and unsuccessful responses, and detect abnormal behaviors \citep{stadler2019taking, ren2019exploring, wang2018detecting}.

Although the methodology development has been prosperous, the software development for analyzing process data has not been done in parallel. Packages and toolboxes for managing and analyzing continuous time series data are abound in literature, but few has been developed for categorical sequence data similar to process data. \pkg{TraMineR} \citep{traminer} is an \proglang{R} \citep{citeR} package designed for describing and analyzing discrete sequence data. However, it cannot be applied directly to process data coming from the log files of computer-based items. Process data are recorded as timestamped action sequences whose lengths vary greatly in practice while the sequences handled by \pkg{TraMineR} consist of states observed at given time points which are the same across observations. It is very difficult to organize process data as dataframes that can be used by \pkg{TraMineR}. Moreover, process data usually have much more possible states (actions) and are much noisier than the sequence data dealt with by \pkg{TraMineR}.
Although neural networks have been shown useful for analyzing process data and more general education data, building a properly working neural network model usually requires a significant amount of effort even with the help of high level neural network application programming interfaces (APIs) such as the \proglang{Python} library \pkg{Keras} and its \proglang{R} interface \pkg{keras} \citep{keras}.
The lack of software for practitioners' usage hinders the wide applications of the state-of-the-art methods and the discovery of scientific results. 


We try to fill in this gap by developing an \proglang{R} package \pkg{ProcData}. This package provides easy-to-use tools for processing, describing, and analyzing process data. It includes functions implementing the state-of-the-art methods for process data analysis. It also provides functions that wrap a series of \pkg{keras} functions to facilitate implementation of neural networks used for analyzing process data.
\pkg{ProcData} is available from the Comprehensive R Archive Network (CRAN) at \url{http://CRAN.R-project.org/package=ProcData} and under development on GitHub at \url{https://github.com/xytangtang/ProcData}.

The remaining sections of this article are organized as follows. We introduce the features of the \pkg{ProcData} package in Section \ref{sec:package}. A case study of the climate control item in PISA 2012 is presented in Section \ref{sec:example} to demonstrate the usage of the package. Summary is included in Section \ref{sec:summary}.



\section[The ProcData Package]{The \pkg{ProcData} Package} \label{sec:package}
\begin{table}[t!]
\centering
\begin{tabular}{cc}
\hline
Topic & Objects\\
\hline
\multirow{2}{*}{Process data} & \code{proc}, \code{print.proc}, \code{summary.proc}\\
 & \code{cc_data}, \code{seq_gen}, \code{seq_gen2}, \code{seq_gen3}$^\dagger$\\
\hline
Process input and output & \code{read.seqs}, \code{write.seqs}\\
\hline
\multirow{2}{*}{Process manipulation} & \code{remove_repeat}, \code{remove_action},\\
 & \code{replace_action}, \code{combine_actions}\\
\hline
\multirow{2}{*}{Feature extraction} & \code{seq2feature_mds}, \code{chooseK_mds}, \\
 & \code{seq2feature_seq2seq}$^\dagger$, \code{chooseK_seq2seq}$^\dagger$\\
\hline
Sequence model & \code{seqm}$^\dagger$, \code{predict.seqm}$^\dagger$ \\
\hline
\end{tabular}
\caption{Summary of \pkg{ProcData} features. The functions marked by $^\dagger$ depend on the \pkg{keras} package.}\label{table:summary}
\end{table}
\pkg{ProcData} provides tools for analyzing process data. It defines an \proglang{S}3 object for process data and includes functions for data processing, exploration, and modeling. The main features of \pkg{ProcData} is summarized in Table~\ref{table:summary}. The details of these features are described in the sequel.

Some features of \pkg{ProcData} require construction and training of neural networks. \pkg{ProcData} relies on \proglang{R} package \pkg{keras} for achieving this functionality. The \pkg{keras} package is an \proglang{R} interface to \proglang{Keras}, which is a high level neural network API developed for fast experimentation of neural networks in \proglang{Python}. The functions in \pkg{ProcData} that are built on functions in \pkg{keras} is marked by $^\dagger$ in Table~\ref{table:summary}. The installation guide of the \pkg{keras} package can be found at \url{https://keras.rstudio.com/}. If \pkg{keras} is not installed properly, calling these marked functions will lead to an error while other functions in \pkg{ProcData} can still be used normally.

\subsection{Datasets and Process Generators}
We include in this package the climate control data in PISA 2012. The dataset contains the response processes and the dichotomous response outcomes of 16,763 students. The item interface is briefly described in the introduction and is available online at \url{http://www.oecd.org/pisa/test-2012/testquestions/question3/}.
The data can be loaded by
\begin{Schunk}
\begin{Sinput}
R> library("ProcData")
R> data("cc_data")
\end{Sinput}
\end{Schunk}
The data object \code{cc_data} is a list of two elements, \code{seqs} and \code{responses}. The response outcomes are contained in \code{responses} as a numeric vector. The response processes are stored in \code{seqs} as an object of class \class{proc}, which will be specified in Section \ref{sec:proc}.

Three functions, \fct{seq\_gen}, \fct{seq\_gen2}, and \fct{seq\_gen3}, are included in this package to simulate process data.
The values of these functions (both action and timestamp sequences) are objects of class \class{proc}. All the three functions generate the inter-arrival time (time elapsed between two consecutive actions) independently from a provided distribution if \code{include_time = TRUE}. 
Each of them has a different underlying model to generate action sequences. 

The values of \fct{seq\_gen} resemble those from an item similar to the climate control item. In particular, participants are asked to answer a question by running simulated experiments in which two conditions can be controlled. A simulated experiment can be run by setting the two conditions at one of the given choices and clicking the ``RUN'' button. An example of the action sequence generated by \fct{seq\_gen} is 
\texttt{``Start, OPT1\_3, OPT2\_2, RUN, OPT1\_1, OPT2\_2, RUN, OPT1\_1, OPT2\_1, RUN, CHECK\_D, End''}.

Function \fct{seq\_gen2} generates action sequences according to a Markov model with a provided transition probability matrix. Given an $N \times N$ transition probability matrix $\mathbf{P} = (p_{ij})$ among $N$ actions, an action sequence is generated by setting $S_1 = a_1$ and the remaining actions are generated according to $P(S_{t+1} = a_j \, | \, S_{t} = a_i) = p_{ij}$ for $t \geq 1$. The action sequence terminates once the predetermined terminating action (one of the $N$ actions) is generated.

Function \fct{seq\_gen3} generates action sequences according to a recurrent neural network (RNN) \citep[Chapter 10]{goodfellow2016deep}. More specifically, an RNN with parameter $\boldsymbol{\eta}$ is used to define the distribution of the action to be taken $S_{t}$ given all previous actions $S_{1}, \ldots, S_{t-1}$. An action sequence is generated recursively in \fct{seq\_gen3} according to the distribution until the predetermined terminating action appears. The parameter vector $\boldsymbol{\eta}$ in the RNN can either be provided by the user or be randomly generated in \fct{seq\_gen3}.

\subsection[S3 class proc]{\proglang{S}3 class \class{proc}}\label{sec:proc}
\pkg{ProcData} defines an \proglang{S}3 class \class{proc} for organizing process data. An object of class \class{proc} is a list of two elements, \code{action_seqs} and \code{time_seqs}. In a \class{proc} object storing response processes $O_1, \ldots, O_n$, \code{action_seqs} is a list with elements $S_1, \ldots, S_n$ and \code{time_seqs} is a list with elements $T_1, \ldots, T_n$.
The names of the elements in \code{action_seqs} and \code{time_seqs} are the identity of the respondents. If the timestamp sequences are not available, \code{time_seqs} is set to \code{NULL}. The \code{seqs} element in \code{cc_data} and the object returned by the three process generators are \class{proc} objects
\begin{Schunk}
\begin{Sinput}
R> class(cc_data$seqs)
\end{Sinput}
\begin{Soutput}
[1] "proc"
\end{Soutput}
\begin{Sinput}
R> seqs <- seq_gen(100)
R> class(seqs)
\end{Sinput}
\begin{Soutput}
[1] "proc"
\end{Soutput}
\end{Schunk}

An object of class \class{proc} is printed by the \code{print} method for class \class{proc}, \fct{print.proc}. Note that the \class{proc} object \code{seqs} below does not contain timestamp sequences.
\begin{Schunk}
\begin{Sinput}
R> seqs
\end{Sinput}
\begin{Soutput}
'proc' object of  100  processes

First  5  processes:

1 
      Step 1 Step 2
Event Start  End   

2 
      Step 1 Step 2
Event Start  End   

3 
      Step 1 Step 2 Step 3 Step 4 Step 5
Event Start  OPT1_2 OPT2_2 RUN    End   

4 
      Step 1 Step 2  Step 3
Event Start  CHECK_B End   

5 
      Step 1 Step 2  Step 3
Event Start  CHECK_C End   
\end{Soutput}
\begin{Sinput}
R> print(cc_data$seqs, index=3)
\end{Sinput}
\begin{Soutput}
'proc' object of  16763  processes


ARE000000300079 
      Step 1 Step 2 Step 3 Step 4 Step 5 Step 6 Step 7 Step 8 Step 9
Event start  1_1_1  reset  0_0_1  reset  0_1_0  reset  1_0_0  end   
Time    0.0  113.2  119.1  122.0  135.4  138.5  147.8  149.8  157.0 
\end{Soutput}
\end{Schunk}

We also extend the \code{summary} method for class \class{proc}. Function \fct{summary.proc} returns a list containing the following components:
\begin{itemize}
\item \code{n_seq}: the number of response processes;
\item \code{n_action}: the number of distinct actions;
\item \code{actions}: $\mathcal{A}$, the set of all possible actions, 
\item \code{seq_length}: a numeric vector of length \code{n_seq} containing the sequence length of each process;
\item \code{action_freq}: a numeric vector of length \code{n_action} containing the action frequencies;
\item \code{action_seqfreq}: a numeric vector of length \code{n_action} recording the number of sequences that each action appears;
\item \code{trans_count}: an \code{n_action}$\times$\code{n_action} matrix with the $ij$th element being the count of bigram ``\code{actions[i]}, \code{actions[j]}'';
\item \code{total_time}: a summary of the response time of the \code{n_seq} response processes;
\item \code{mean_react_time}: a summary of the mean reaction time (response time divided by sequence length) of the \code{n_seq} response processes.
\end{itemize}

\subsection{Read and Write Process Data}
\begin{figure}[htb]
\centering
\includegraphics{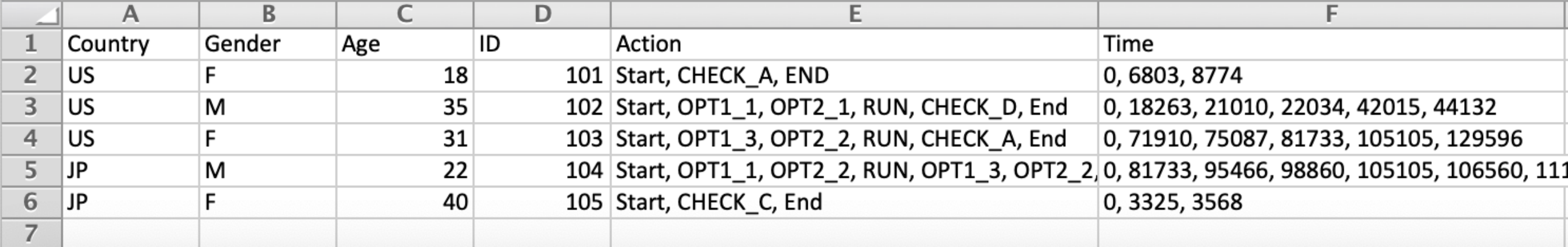}
\caption{\label{fig:style_single} Screenshot of a CSV file storing response processes in ``single'' style.}
\end{figure}

\begin{figure}[htb]
\centering
\includegraphics{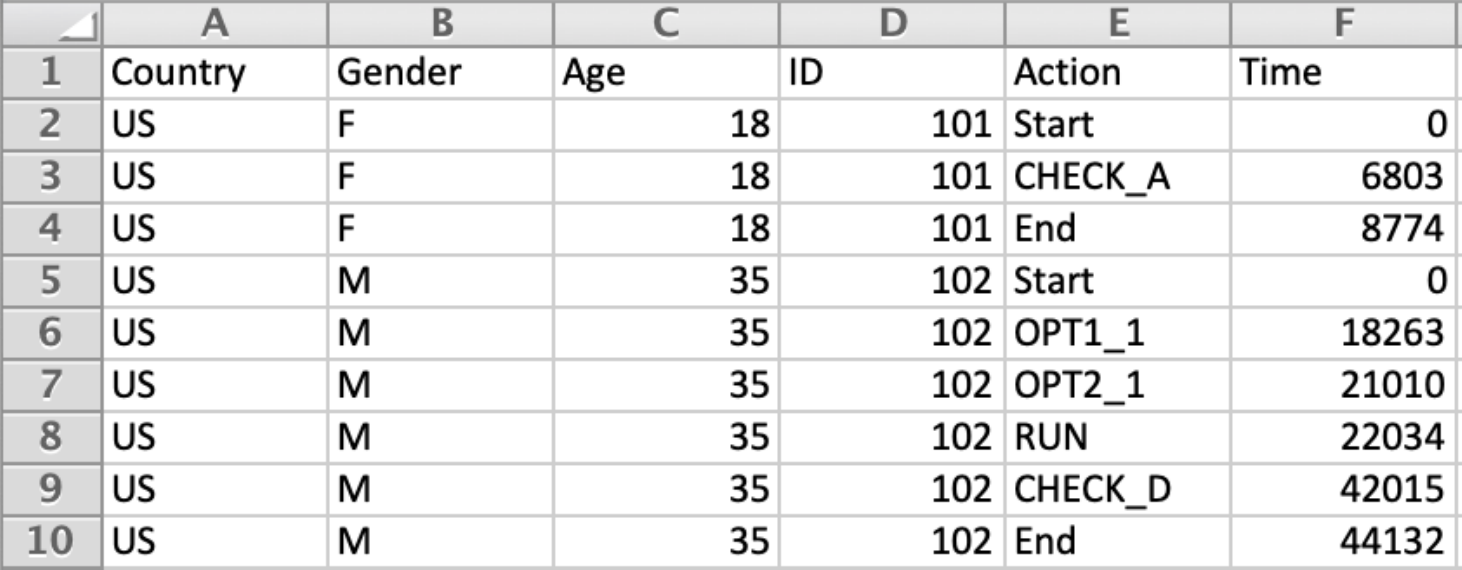}
\caption{\label{fig:style_multiple} Screenshot of a CSV file storing response processes in ``multiple'' style.}
\end{figure}

Process data are usually stored as comma separated values (CSV) files. \pkg{ProcData} provides functions \fct{read.seqs} and \fct{write.seqs} to read process data from a CSV file as a \class{proc} object and to write a \class{proc} object to a CSV file. 
The functions accommodate CSV files of two styles, ``single'' and ``multiple''. In both styles, all the action sequences form a column (action column) and all the timestamp sequences form another column (time column).
In the ``single'' style, the process for one respondent takes up one row in the CSV file. The entire action sequence of the respondent is stored as one entry of the action column. The timestamp sequence of the respondent is stored in the corresponding entry of the time column. Figure \ref{fig:style_single} presents a CSV file storing five response processes in the ``single'' style.
In the ``multiple'' style, each action in the process and its timestamp occupy one row in the CSV file. A process of length $L$ takes up $L$ consecutive rows. A CSV file that stores two response processes in the ``multiple'' style is displayed in Figure \ref{fig:style_multiple}. The two response processes are the same as the first two shown in Figure \ref{fig:style_single}. In \fct{read.seqs} and \fct{write.seqs}, both styles are accommodated. The style can be specified by setting the argument \code{style}.
\begin{Schunk}
\begin{Sinput}
R> write.seqs(cc_data$seqs, file="seqs_format_multiple.csv", 
+             style = "multiple", id_var="ID", action_var="Action", 
+             time_var="Time")
R> write.seqs(cc_data$seqs, file="seqs_format_single.csv", 
+             style = "single", id_var="ID", action_var="Action", 
+             time_var="Time", step_sep=",")
R> seqs_multiple <- read.seqs("seqs_format_multiple.csv", 
+                             style = "multiple", id_var = "ID", 
+                             action_var = "Action", time_var = "Time")
R> seqs_single <- read.seqs("seqs_format_single.csv", style = "single", 
+                           id_var = "ID", action_var = "Action", 
+                           time_var = "Time", step_sep = ",")
R> all.equal(seqs_multiple, seqs_single)
\end{Sinput}
\begin{Soutput}
[1] TRUE
\end{Soutput}
\end{Schunk}

\subsection{Sequence Manipulation}
\pkg{ProcData} provides functions for presponse process manipulation. These functions can be used for process data cleaning. Function \fct{sub\_seqs} subsets a set of response processes. 
\begin{Schunk}
\begin{Sinput}
R> seqs <- sub_seqs(cc_data$seqs, 4*1:10)
R> print(seqs, 1)
\end{Sinput}
\begin{Soutput}
'proc' object of  10  processes

First  1  processes:

ARE000000400093 
      Step 1 Step 2 Step 3 Step 4 Step 5 Step 6 Step 7 Step 8 Step 9
Event start  1_0_0  0_0_0  reset  0_0_0  0_0_1  0_0_2  reset  0_1_0 
Time    0.0   36.9   41.2   43.3   44.5   50.5   61.9   66.1   68.7 
      Step 10 Step 11 Step 12 Step 13 Step 14 Step 15 Step 16 Step 17
Event 0_2_0   -1_2_0  reset   2_0_0   reset   2_0_0   2_0_0   2_0_0  
Time   72.3    92.1    94.5    97.2   110.3   112.9   116.5   117.8  
      Step 18 Step 19 Step 20 Step 21 Step 22 Step 23 Step 24 Step 25
Event reset   0_2_0   0_2_0   0_2_0   reset   0_0_2   0_0_2   0_0_2  
Time  121.1   123.6   124.8   125.6   127.5   129.4   130.0   130.7  
      Step 26 Step 27 Step 28
Event reset   1_1_1   end    
Time  131.7   147.8   171.6  
\end{Soutput}
\end{Schunk}
Function \fct{remove\_repeat} removes the consecutive repeated actions and their timestamps.
\begin{Schunk}
\begin{Sinput}
R> seqs1 <- remove_repeat(seqs)
R> print(seqs1, 1)
\end{Sinput}
\begin{Soutput}
'proc' object of  10  processes

First  1  processes:

ARE000000400093 
      Step 1 Step 2 Step 3 Step 4 Step 5 Step 6 Step 7 Step 8 Step 9
Event start  1_0_0  0_0_0  reset  0_0_0  0_0_1  0_0_2  reset  0_1_0 
Time    0.0   36.9   41.2   43.3   44.5   50.5   61.9   66.1   68.7 
      Step 10 Step 11 Step 12 Step 13 Step 14 Step 15 Step 16 Step 17
Event 0_2_0   -1_2_0  reset   2_0_0   reset   2_0_0   reset   0_2_0  
Time   72.3    92.1    94.5    97.2   110.3   112.9   121.1   123.6  
      Step 18 Step 19 Step 20 Step 21 Step 22
Event reset   0_0_2   reset   1_1_1   end    
Time  127.5   129.4   131.7   147.8   171.6  
\end{Soutput}
\end{Schunk}
Function \fct{remove\_action} removes a particular set of actions and their timestamps. 
\begin{Schunk}
\begin{Sinput}
R> seqs2 <- remove_action(seqs1, "0_0_0")
R> print(seqs2, 1)
\end{Sinput}
\begin{Soutput}
'proc' object of  10  processes

First  1  processes:

ARE000000400093 
      Step 1 Step 2 Step 3 Step 4 Step 5 Step 6 Step 7 Step 8 Step 9
Event start  1_0_0  reset  0_0_1  0_0_2  reset  0_1_0  0_2_0  -1_2_0
Time    0.0   36.9   43.3   50.5   61.9   66.1   68.7   72.3   92.1 
      Step 10 Step 11 Step 12 Step 13 Step 14 Step 15 Step 16 Step 17
Event reset   2_0_0   reset   2_0_0   reset   0_2_0   reset   0_0_2  
Time   94.5    97.2   110.3   112.9   121.1   123.6   127.5   129.4  
      Step 18 Step 19 Step 20
Event reset   1_1_1   end    
Time  131.7   147.8   171.6  
\end{Soutput}
\end{Schunk}
Function \fct{replace\_action} renames an action.
\begin{Schunk}
\begin{Sinput}
R> seqs3 <- replace_action(seqs2, "reset", "RESET")
R> print(seqs3, 1)
\end{Sinput}
\begin{Soutput}
'proc' object of  10  processes

First  1  processes:

ARE000000400093 
      Step 1 Step 2 Step 3 Step 4 Step 5 Step 6 Step 7 Step 8 Step 9
Event start  1_0_0  RESET  0_0_1  0_0_2  RESET  0_1_0  0_2_0  -1_2_0
Time    0.0   36.9   43.3   50.5   61.9   66.1   68.7   72.3   92.1 
      Step 10 Step 11 Step 12 Step 13 Step 14 Step 15 Step 16 Step 17
Event RESET   2_0_0   RESET   2_0_0   RESET   0_2_0   RESET   0_0_2  
Time   94.5    97.2   110.3   112.9   121.1   123.6   127.5   129.4  
      Step 18 Step 19 Step 20
Event RESET   1_1_1   end    
Time  131.7   147.8   171.6  
\end{Soutput}
\end{Schunk}

Function \fct{combine\_actions} combines a given pattern of consecutive actions into a single action.
\begin{Schunk}
\begin{Sinput}
R> seqs4 <- combine_actions(seqs2, c("0_0_1", "0_0_2"), "BOTTOM_MOVE_ONE")
R> print(seqs4, 1)
\end{Sinput}
\begin{Soutput}
'proc' object of  10  processes

First  1  processes:

ARE000000400093 
      Step 1 Step 2 Step 3 Step 4          Step 5 Step 6 Step 7
Event start  1_0_0  reset  BOTTOM_MOVE_ONE reset  0_1_0  0_2_0 
Time    0.0   36.9   43.3   50.5            66.1   68.7   72.3 
      Step 8 Step 9 Step 10 Step 11 Step 12 Step 13 Step 14 Step 15
Event -1_2_0 reset  2_0_0   reset   2_0_0   reset   0_2_0   reset  
Time   92.1   94.5   97.2   110.3   112.9   121.1   123.6   127.5  
      Step 16 Step 17 Step 18 Step 19
Event 0_0_2   reset   1_1_1   end    
Time  129.4   131.7   147.8   171.6  
\end{Soutput}
\end{Schunk}

\subsection{Feature Extraction}

A major technical difficulty for process data analysis is that response processes are not readily and cannot be easily organized as a matrix. The nonstandard format prevents the direct application of many traditional statistical methods to process data. Feature extraction methods have been shown useful for exploratory process data analysis. These methods circumvent the nonstandard format difficulty by compressing response processes into fixed-dimension vectors that can be easily incorporated in traditional statistical models such as regression models. 
Two feature extraction methods, one based on multidimensional scaling \citep{tang2019mds} and the other based on sequence-to-sequence autoencoder \citep{tang2019seq2seq}, have been proposed for response processes. Both methods are implemented in \pkg{ProcData}.

\subsubsection{Feature Extraction via Multidimensional Scaling}
Multidimensional scaling (MDS) \citep{borg2005modern} is a technique that is often used for dimension reduction and data visualization. Its goal is to embed objects in a space in such a way that the dissimilarity between objects is approximated by the distance between their embedded features. Similar objects are located close together while dissimilar objects are far apart.
Given a dissimilarity measure that comprehensively characterizes the difference between objects, the object coordinates obtained from MDS can be viewed as features describing the latent attributes of the objects. Given a dissimilarity measure $d$ and the latent feature dimension $K$, features of a set of response processes $O_1, \ldots, O_n$ are a solution to the optimization problem
\begin{equation}\label{eq:mds_obj}
\min_{\boldsymbol{\theta}_1, \ldots, \boldsymbol{\theta}_n \in \mathbb{R}^K}\sum_{1 \leq i < j \leq n}(d_{ij} - \| \boldsymbol{\theta}_i - \boldsymbol{\theta}_j\|)^2,
\end{equation}
where $d_{ij} = d(O_i, O_j)$ is the dissimilarity between response processes $O_i$ and $O_j$, $\boldsymbol{\theta}_i$ is the feature vector of response process $O_i$, and $\| \mathbf{x} \| = \sqrt{\mathbf{x}^\top \mathbf{x}}$. 

The function that performs MDS feature extraction in \pkg{ProcData} is \fct{seq2feature\_mds}. It takes a \class{proc} object (\code{seqs =}), the number of features to be extracted (\code{K =}), and some other control arguments to calculate the dissimilarity matrix for the input response processes and then performs MDS. 
\begin{Schunk}
\begin{Sinput}
R> seqs <- seq_gen(100)
R> mds_res <- seq2feature_mds(seqs = seqs, K = 10)
R> str(mds_res)
\end{Sinput}
\begin{Soutput}
List of 1
 $ theta: num [1:100, 1:10] -0.195 -0.195 0.272 -0.162 -0.167 ...
  ..- attr(*, "dimnames")=List of 2
  .. ..$ : NULL
  .. ..$ : NULL
\end{Soutput}
\end{Schunk}
Classical MDS (implemented as \fct{cmdscale} in \proglang{R}) approximates the solution to \eqref{eq:mds_obj} by performing eigenvalue decomposition of the matrix $-\frac{1}{2}\mathbf{J}\mathbf{D}^{(2)}\mathbf{J}$ where $\mathbf{J} = \mathbf{I} - \frac{1}{n}\mathbf{1}^\top\mathbf{1}$ and $\mathbf{D}^{(2)} = (d_{ij}^2)$. The computational complexity of the algorithm is $O(n^3)$, which is very expensive if the number of processes $n$ is large. Moreover, the $n \times n$ dissimilarity matrix $D$ consumes a large amount of memory. 
To accomodata a large $n$, the algorithm proposed in \citet{paradis2018multidimensional} is implemented in \fct{seq2feature\_mds}. This algorithm first chooses a small subset $\Omega$ of the objects and obtains $\hat{\boldsymbol{\theta}}_i, i \in \Omega$ by performing classical MDS on this subset. Then it minimizes 
\begin{equation}
F(\boldsymbol{\theta}_i) = \sum_{j \in \Omega} (d_{ij} - \| \boldsymbol{\theta}_i - \hat{\boldsymbol{\theta}}_j\|)^2
\end{equation}
by the BFGS method \citep{broyden1970convergence,fletcher1970new,goldfarb1970family,shanno1970conditioning} for each $i \not\in \Omega$. In this way, only the dissimilarities for $O(mn)$ pairs of objects are calculated where $m$ is the subset size and the eigenvalue decomposition for a large matrix is avoided. 

In \fct{seq2feature\_mds}, argument \code{method} specifies the algorithm for the feature extraction. If \code{method = "small"}, \fct{cmdscale} is called to perform classical MDS. If \code{method = "large"}, the algorithm for large datasets is used. By default (\code{method = "auto"}), \fct{seq2feature\_mds} selects the method according to the sample size.

Two choices of dissimilarity measures are implemented in \fct{seq2feature\_mds}. One is the optimal symbol similarity (OSS) measure (\code{dist_type = "oss_action"}) proposed in \cite{gomez2008similarity}. It takes into account the difference between two action sequences $S_i$ and $S_j$. The other choice (\code{dist_type = "oss_both"}) is a time-weighted version of the OSS measure. It takes the difference in both action sequences and the timestamp sequences into consideration. If a dissimilarity measure other than the two choices are desired, a dissimilarity matrix can be pre-computed and passed to \fct{seq2feature\_mds} via \code{seqs}.

Following the recommendation in \citet{tang2019mds}, \fct{seq2feature\_mds} performs principal component analysis on the extracted features to enhance the interpretability of the features. This step can be turned off by setting \code{pca = FALSE}. By default, \fct{seq2feature\_mds} returns a list containing the latent features (\code{theta}) and the value of the optimized objective function (\code{loss}). The dissimilarity matrix (\code{dist_mat}) can be returned by setting \code{return_dist = TRUE}.

\pkg{ProcData} provides a function \fct{chooseK\_mds} to select the latent feature dimension ($K$) by $k$-fold cross-validation. The basic usage is demonstrated in the example below.
\begin{Schunk}
\begin{Sinput}
R> seqs <- seq_gen(100)
R> cv_res <- chooseK_mds(seqs = seqs, K_cand = 5:10, n_fold = 5, 
+                        return_dist = TRUE)
R> str(cv_res)
\end{Sinput}
\begin{Soutput}
List of 4
 $ K       : int 8
 $ K_cand  : int [1:6] 5 6 7 8 9 10
 $ cv_loss : num [1:6, 1] 0.00399 0.00296 0.00265 0.00239 0.00242 ...
 $ dist_mat: num [1:100, 1:100] 0 0 0.514 0.267 0.267 ...
\end{Soutput}
\begin{Sinput}
R> mds_res <- seq2feature_mds(seqs = cv_res$dist_mat, K = cv_res$K)
R> str(mds_res)
\end{Sinput}
\begin{Soutput}
List of 1
 $ theta: num [1:100, 1:8] -0.195 -0.195 0.272 -0.162 -0.167 ...
  ..- attr(*, "dimnames")=List of 2
  .. ..$ : NULL
  .. ..$ : NULL
\end{Soutput}
\end{Schunk}

\subsubsection{Sequence-to-Sequence Autoencoder Feature Extraction}
\begin{figure}[htb]
\centering
\includegraphics[width=0.65\textwidth]{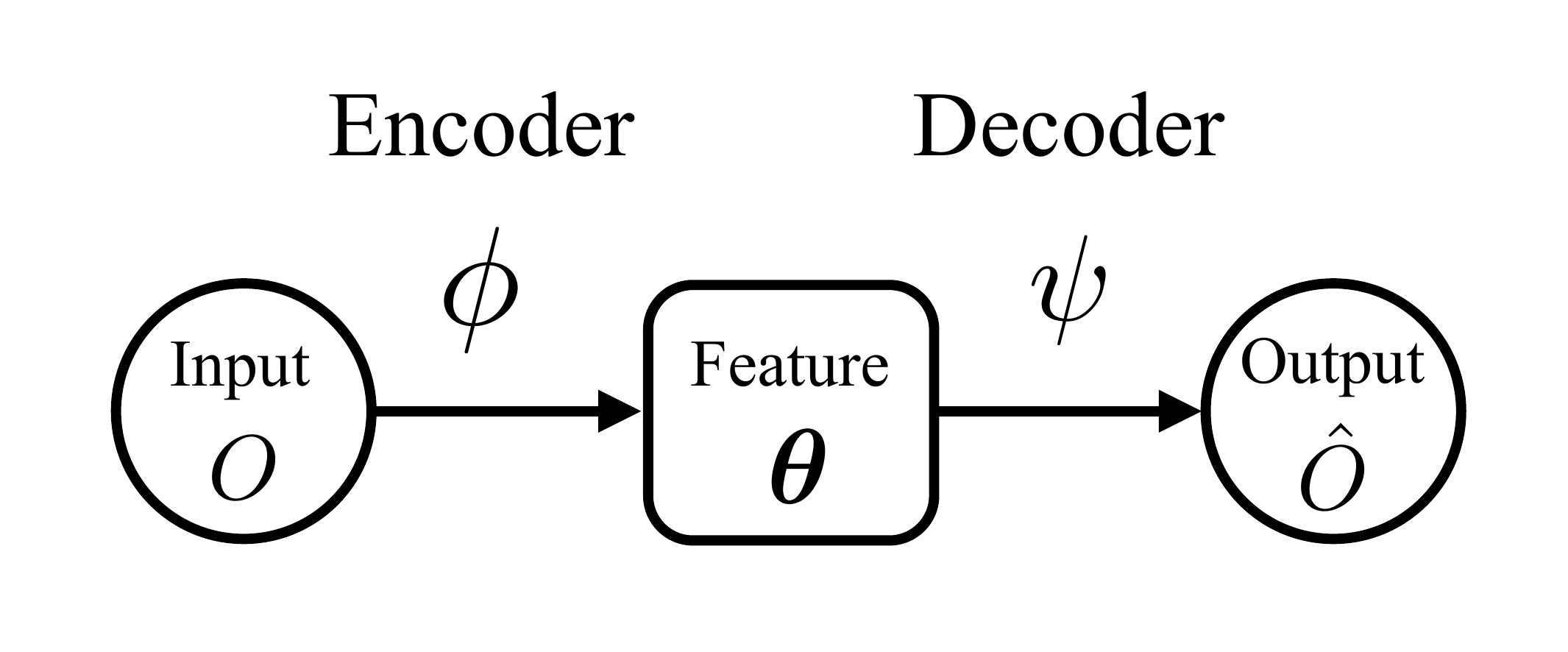}
\caption{The structure of autoencoders.} \label{fig:ae}
\end{figure}

Autoencoder \citep[Chapter 14]{goodfellow2016deep} is a class of neural networks aiming at reconstructing the input in the output. It consists of two components (Figure \ref{fig:ae}), an encoder that maps the complex and/or high dimensional input to a low dimensional vector and a decoder that reconstructs the input from the low dimensional vector. The low-dimensional vector produced by the encoder contains condensed information to rebuild the input. Therefore, its elements can be treated as features of the input data. 

In \pkg{ProcData}, \fct{seq2feature\_seq2seq} extracts $K$ (\code{K =}) features from a given set of response processes (\code{seqs =}) by fitting a sequence autoencoder (SeqAE). The SeqAEs constructed in \fct{seq2feature\_seq2seq} can be classified into action SeqAE, time SeqAE, and action-time SeqAE depending on the input process.
As the names suggest, an action SeqAE only deals with action sequences, a time SeqAE only deals with timestamp sequences, and an action-time SeqAE deals with both action and timestamp sequences in response processes. The desired type of SeqAE can be specified through argument \code{ae_type} in the function.

\begin{figure}[htb]
\centering
\includegraphics[width=\textwidth]{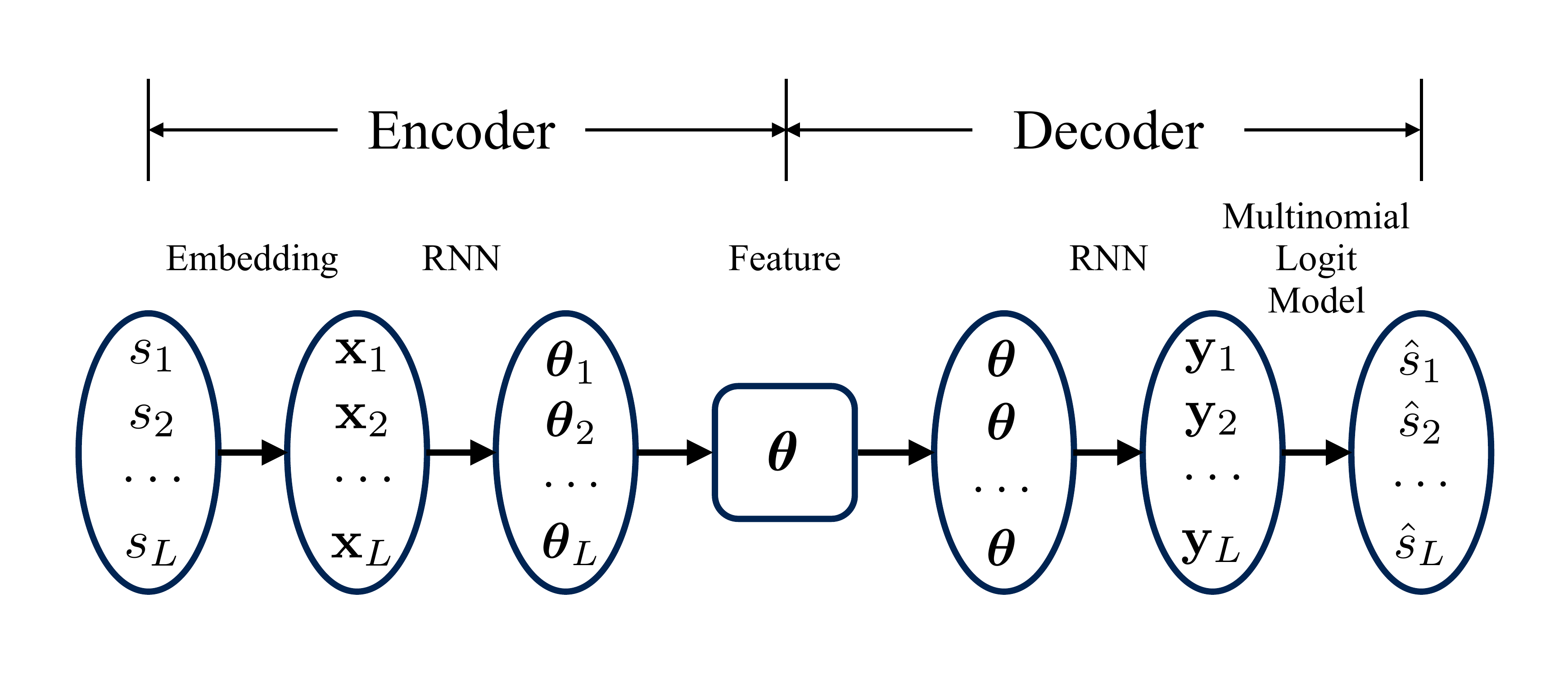}
\caption{The structure of action seqeunce autoencoders.}\label{fig:action_ae}
\end{figure}

The structure of an action SeqAE is depicted in Figure \ref{fig:action_ae}. The encoder of an action SeqAE consists of three steps. In the first step, each unique action in $\mathcal{A}$ is associated with a numeric vector (embedding) in $\mathbb{R}^K$ so that an input action sequence is transformed into a sequence of $K$-dimensional embeddings. In the second step, an RNN is used to sequentially summarize the information in the embedding sequence up to a time step. In the last step, a feature vector is constructed from the output vectors of the encoder RNN.
The decoder of an action SeqAE also consists of three steps. In the first step, the feature vector is repeated to form a sequence of vectors which are then passed into another RNN in the second step to obtain another sequence of vectors. Each vector in this latter sequence contains the information of the action at the corresponding time step. In the last step, a fully connected layer with a softmax activation (multinomial logistic model) is used to construct a probability distribution $\hat{s}_l = (\hat{s}_{l1}, \hat{s}_{l2}, \ldots, \hat{s}_{lN})$ on $\mathcal{A}$ at each time step $l$ from the corresponding vector obtained from the decoder RNN.

\begin{figure}[htb]
\centering
\includegraphics[width=\textwidth]{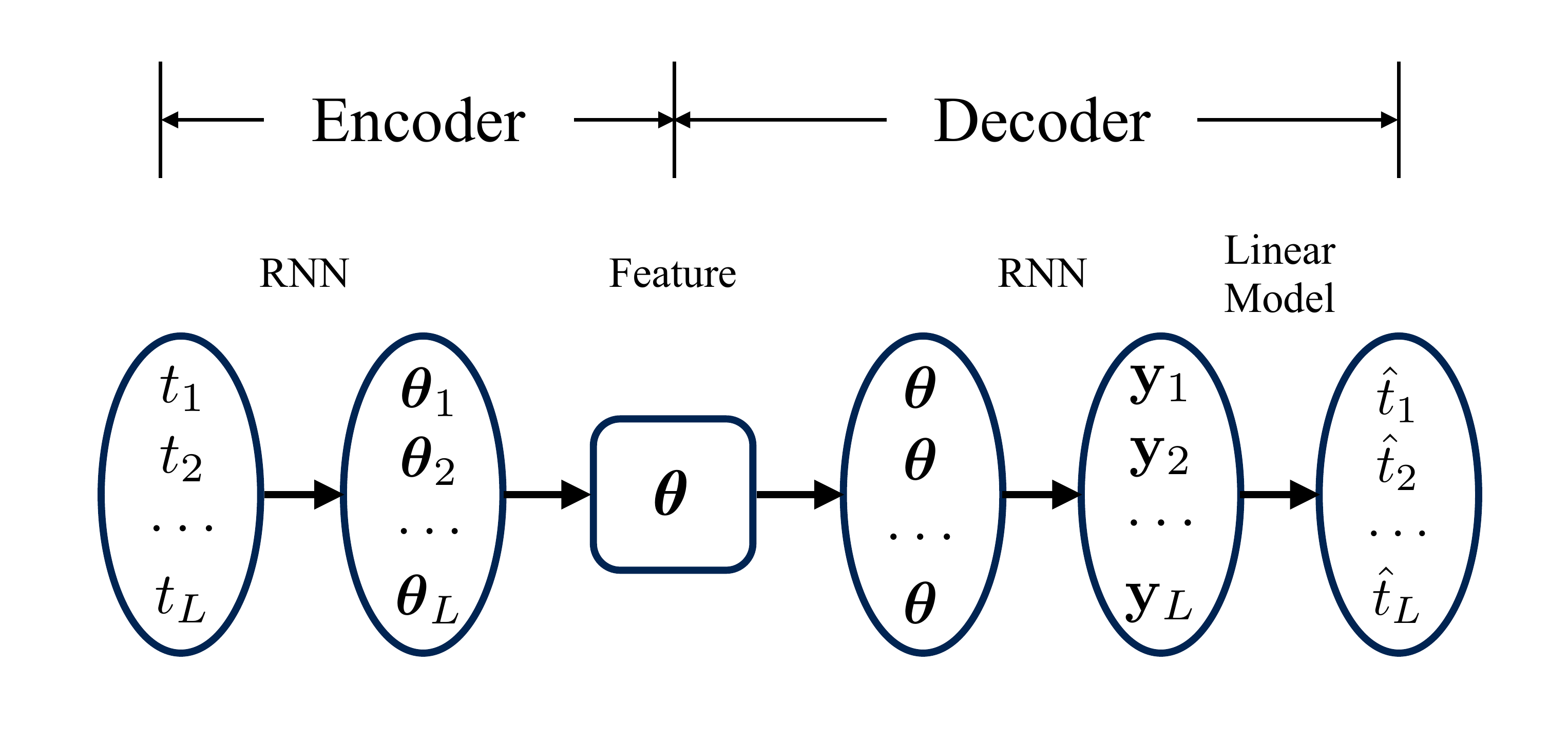}
\caption{The structure of time sequence autoencoders.}\label{fig:time_ae}
\end{figure}

The structure of a time SeqAE (Figure \ref{fig:time_ae}) is similar to that of an action SeqAE. RNNs are used in both the encoder and the decoder to summarize the sequential information. The feature vector is obtained from the output vectors of the encoder RNN. However, since timesteps are numeric, embedding is not needed in the encoder. Also, in the last step of the decoder, the timesteps are reconstructed through a fully connected layer with a ReLU activation instead of a softmax activation.

\begin{figure}[htb]
\centering
\includegraphics[width=\textwidth]{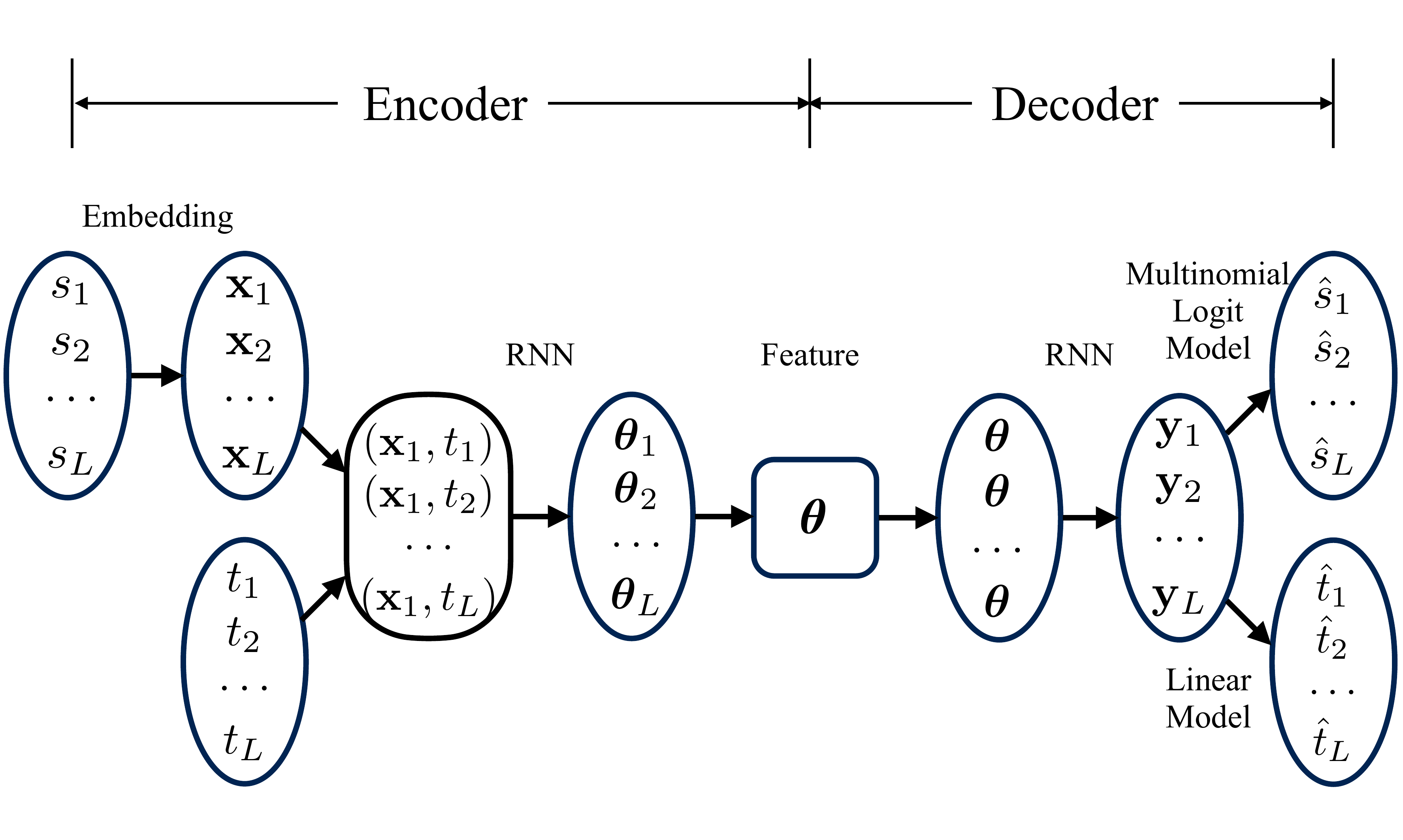}
\caption{The structure of action-time autoencoders.}\label{fig:action_time_ae}
\end{figure}

The structure of an action-time SeqAE is a combination of an action SeqAE and a time SeqAE as shown in Figure \ref{fig:action_time_ae}. Given a response process of length $L$, the encoder of an action-time SeqAE first transforms the action sequence into a $K$-dimensional embedding sequence. The embedding sequence is then combined with the timestamp sequence to form a $(K+1)$-dimensional sequence that serves as the input of the encoder RNN. The encoder RNN produces a sequence of $L$ output vectors, each of which is $K$-dimensional. Tthe feature vector is then computed from the output of the encoder RNN. 
The first two steps of the decoder of an action-time SeqAE is the same as that of an action SeqAE. Once the output vectors from the decoder RNN are obtained, they are passed into a fully connected layer with a softmax activation to construct a probability distribution on $\mathcal{A}$ and into another fully connected layer but with a ReLU activation to reproduce the timestamps.

Two methods can be used to construct the feature vector from the output vectors of the encoder RNN for all three types of SeqAEs. In the first method (\code{method = "last"}), the feature vector is the last output vector of the decoder RNN. In the other method (\code{method = "avg"}), the feature vector is the average of the $L$ output vectors.

By default, the original timestamp sequences are used in time SeqAE and action-time SeqAE. The inter-arrival time sequences can be used in replace of the timestamp sequences by setting \code{cumulative = FALSE}. The natural logorithm of the timestamps or the inter-arrival time will be used if \code{log = TRUE}. There are two choices of the recurrent units in the encoder and decoder RNNs, long-short-memory (LSTM) unit \citep{hochreiter1997long} (\code{rnn_type = "lstm"}) and the gated recurrent unit (GRU) \citep{cho2014learning} (\code{rnn_type = "gru"}).

Once the structure of a SeqAE is selected, the parameters of the autoencoder, including the embeddings and the parameters in the encoder RNN, the decoder RNN, and the fully connected layers, are estimated by minimizing a discrepancy between the output and the input over a set of response processes. To be more specific, the estimated parameter vector $\hat{\boldsymbol{\eta}}$ is obtained by minimizing 
\begin{equation}\label{eq:obj_ae}
F(\bm \eta) = \sum_{i \in \Omega} \delta({O}_i, \hat{O}_i)
\end{equation}
where $\hat{O}$ denote the output of a SeqAE for input $O$ and $\delta (O, \hat{O})$ is a function measuring the discrepancy between $O$ and $\hat{O}$. For an action SeqAE, 
$$\delta(O, \hat{O}) = \delta_{\text{a}}(S, \hat{S}) = -\frac{1}{L}\sum_{l=1}^{L} \sum_{j=1}^N \mathbf{1}\{s_l = a_j\} \log(\hat{s}_{lj}).$$
For a time SeqAE, $$\delta(O, \hat{O}) = \delta_{\text{t}}(T, \hat{T}) = \sum_{l=1}^{L} (t_l - \hat{t}_l)^2.$$ 
For an action-time SeqAE, $$\delta(O, \hat{O}) = w_a \delta_{\text{a}}(S, \hat{S}) + w_t \delta_{\text{t}}(T, \hat{T}),$$ where $\mathbf{w} = (w_a, w_t)$, specified through argument \code{weights}, assigns different weights on the discrepancy between action sequences and the discrepancy between time sequences.

In \fct{seq2feature\_seq2seq}, the objective function \eqref{eq:obj_ae} is minimized by stochastic approximation. The available optimizers are stochastic gradient descent (\code{optimizer_name = "sgd"}) \citep{robbins1951stochastic}, Adam (\code{optimizer_name = "adam"}) \citep{kingma2015adam}, AdaDelta (\code{optimizer_name = "adadelta"}) \citep{zeiler2012adadelta}, and RMSprop (\code{optimizer_name = "rmsprop"}) \citep{rmsprop}. The (baseline) step size used in the optimzer is supplied through argument \code{step_size}. The training and validation sets are specified in \code{samples_train} and \code{samples_valid}, respectively. To avoid overfitting, the optimization algorithm is run for a fixed number of epochs (\code{n_epoch =}). At the end of each epoch, the autoencoder is evaluated on the validation set. The parameter value that produces the lowest loss on the validation set is used to produce extracted features. 

Similar to \code{seq2feature\_mds}, principal features, namely, the principal components of the raw features extracted from a SeqAE, are output by default. This step can be turned off by setting \code{pca = FALSE}.

The construction of the architecture of SeqAEs and the training procedures in \fct{seq2feature\_seq2seq} are achieved by calling appropriate functions in package \pkg{keras}. If \pkg{keras} is not installed properly, an error will occur.

\pkg{ProcData} also provides function \fct{chooseK\_seq2seq} to choose the latent feature dimension by cross-validation. The example below demonstrates the basic usage of \fct{seq2feature\_seq2seq} and \fct{chooseK\_seq2seq}. 
\begin{Schunk}
\begin{Sinput}
R> seqs <- seq_gen(100)
R> cv_res <- chooseK_seq2seq(seqs = seqs, ae_type = "action", 
+                            K_cand = c(5, 10), n_epoch = 10, 
+                            verbose = FALSE)
R> str(cv_res)
\end{Sinput}
\begin{Soutput}
List of 3
 $ K      : num 10
 $ K_cand : num [1:2] 5 10
 $ cv_loss: num [1:2, 1] 12 11.4
\end{Soutput}
\begin{Sinput}
R> samples_train <- sample(1:100, 80)
R> samples_valid <- setdiff(1:100, samples_train)
R> seq2seq_res <- seq2feature_seq2seq(seqs = seqs, ae_type = "action", 
+                                     K = cv_res$K, 
+                                     samples_train = samples_train, 
+                                     samples_valid = samples_valid, 
+                                     verbose = FALSE)
R> str(seq2seq_res)
\end{Sinput}
\begin{Soutput}
List of 3
 $ train_loss: num [1:50] 2.48 2.47 2.46 2.44 2.42 ...
 $ valid_loss: num [1:50] 2.47 2.46 2.45 2.44 2.42 ...
 $ theta     : num [1:100, 1:10] -3.21 -3.21 -3.52 -4.21 -3.72 ...
  ..- attr(*, "dimnames")=List of 2
  .. ..$ : NULL
  .. ..$ : chr [1:10] "PC1" "PC2" "PC3" "PC4" ...
\end{Soutput}
\end{Schunk}

\subsection{Feature extraction with targeted variable}
\begin{figure}[htb]
\includegraphics[width=\textwidth]{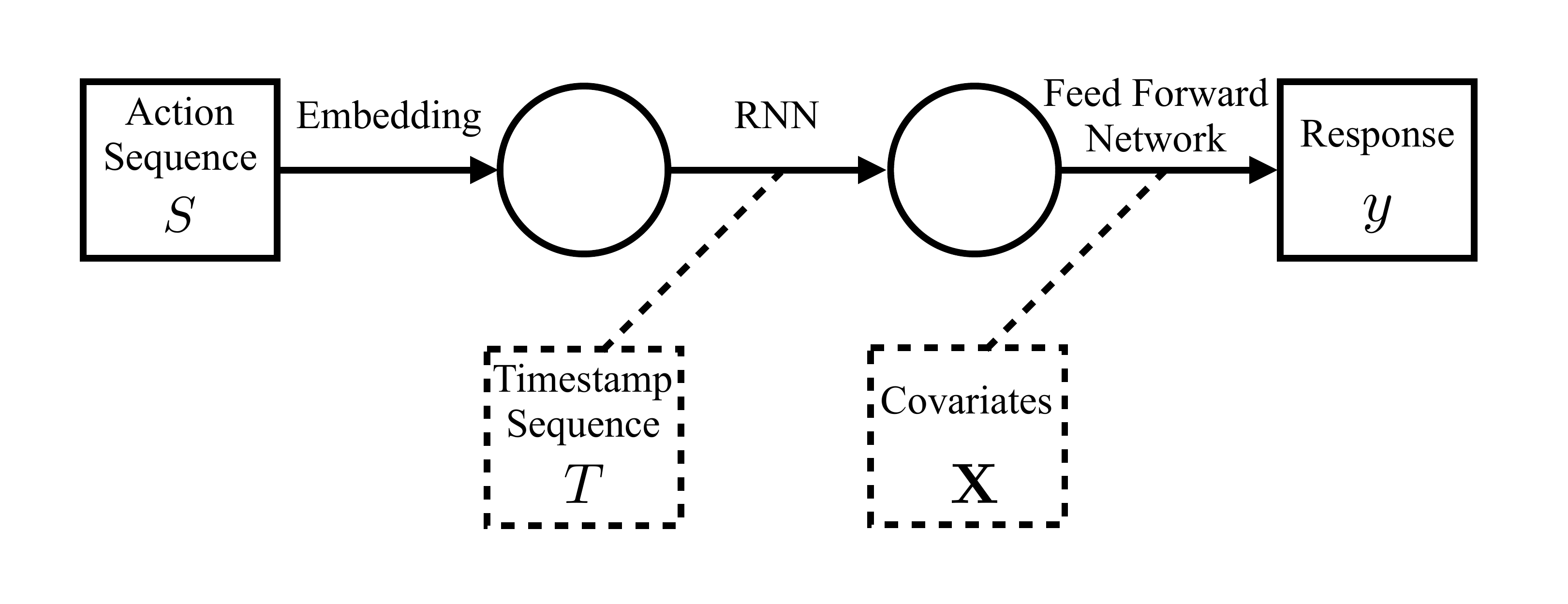}
\caption{The structure of sequence models. The dashed squares represent optional inputs.}\label{fig:seqm}
\end{figure}

The previous feature extraction methods do not have a particular target prediction variable. The features mainly capture variations among response processes. We further present a function that extracts features targeting a particular variable.
Function \code{seqm} fits a sequence model that relates a response process (\code{seqs =}) and covariates (\code{X =}) with a binary (\code{response_type = "binary"}) or numeric (\code{response_type = "scale"}) response variable (\code{response =}).
A sequence model is essentially a neural network, whose architecture is summarized in Figure \ref{fig:seqm}. Given a response process, a sequence model built by \fct{seqm} first transforms the action sequence into a sequence of $K$-dimensional numeric vectors through an embedding step. The dimension of the embeddings is specified through \code{K_emb}. The embedding sequence is then fed into an RNN to process the information sequentially. The dimension of the output vectors of the RNN is specified through \code{K_rnn}. The last output vector from the RNN is used as the input of a single-layer or multi-layer feedforward neural network whose output is the target response variable. To construct a feedforward neural network with more than one layers, set \code{n_hidden} as the number of intermediate layers (the total number of layers minus one) and specify the dimensions of the intermediate layers as a vector through \code{K_hidden}.

If \code{include_time = TRUE} and timestamp sequences are available in the reponse processes, the embedding sequence is combined with the timestamp sequence or the inter-arrival time sequence (\code{time_interval = TRUE}) to form a sequence of $K+1$-dimensional numeric vectors before fed into the RNN. The logorithms of the time-related sequences are used by default. To use the time-related sequences in their original scale, set \code{log_time = FALSE}.

If covariates are provided (\code{X =}), the covariate vector is concatenated to the last output from the RNN. The resulting vector is then used as the input of the feed-forward neural network.

Similar to \fct{seq2feature\_seq2seq}, the parameters in the sequence model are estimated by stochastic approximation. The optimizer, (baseline) step size and the number of epochs to be run is specified through arguments \code{optimizer_name}, \code{step_size}, and \code{n_epoch}, respectively. The training and validation sets are specified by providing either the indices of validation samples or the proportion of the validation samples through \code{index_valid}. In the latter case, the validation samples are randomly selected.

Function \fct{seqm} returns an object of class \class{seqm}, which is a list containing the neural network architecture, the estimated parameters, and other information about modeling fitting. The key elements of a \class{seqm} object are
\begin{itemize}
\item \code{structure}: a character string describing the neural network structure;
\item \code{coefficients}: a list containing estimated parameters;
\item \code{history}: a matrix with two columns giving the training (column 1) and validation (column 2) loss at the end of each epoch.
\end{itemize}

Once a sequence model is fit, prediction can be made by \fct{predict.seqm}. The inputs of \fct{predict.seqm} are a \class{seqm} object returned by \fct{seqm} (\code{object =}), a \class{proc} object containing a set of new response processes (\code{new_seqs =}), and a covariate matrix (\code{new_X =}) of the new response processes. Predictions are produced by evaluating the fitted sequence model at the new response processes and covariates. The outputs of \fct{predict.seqm} are the probabilities of the response being one for binary responses and the expected value of the response for numeric responses. The code below exemplifies the prediction of the binary final response on the climate control item using the problem-solving processes.

\begin{Schunk}
\begin{Sinput}
R> n <- 100
R> data(cc_data)
R> samples <- sample(1:length(cc_data$responses), n)
R> seqs <- sub_seqs(cc_data$seqs, samples)
R> y <- cc_data$responses[samples]
R> x <- matrix(rnorm(n*2), ncol=2)
R> index_test <- 91:100
R> index_train <- 1:90
R> seqs_train <- sub_seqs(seqs, index_train)
R> seqs_test <- sub_seqs(seqs, index_test)
R> actions <- unique(unlist(seqs$action_seqs))
R> ## no covariate is used
R> res1 <- seqm(seqs = seqs_train, response = y[index_train], 
+                response_type = "binary", actions=actions, K_emb = 5, K_rnn = 5, 
+                n_epoch = 20)
R> pred_res1 <- predict(res1, new_seqs = seqs_test)
R> ## use more hidden layers
R> res2 <- seqm(seqs = seqs_train, response = y[index_train],
+               response_type = "binary", actions=actions, K_emb = 5, K_rnn = 5,
+               n_hidden=2, K_hidden=c(10,5), n_epoch = 20)
R> pred_res2 <- predict(res2, new_seqs = seqs_test)
R> ## add covariates
R> res3 <- seqm(seqs = seqs_train, response = y[index_train],
+               covariates = x[index_train, ],
+               response_type = "binary", actions=actions,
+               K_emb = 5, K_rnn = 5, n_epoch = 20)
R> pred_res3 <- predict(res3, new_seqs = seqs_test,
+                       new_covariates=x[index_test, ])
R> ## include time sequences
R> res4 <- seqm(seqs = seqs_train, response = y[index_train],
+               response_type = "binary", actions=actions,
+               include_time=TRUE, K_emb=5, K_rnn=5, n_epoch=20)
R> pred_res4 <- predict(res4, new_seqs = seqs_test)
\end{Sinput}
\end{Schunk}

\section{Examples} \label{sec:example}
In this section, we demonstrate the \pkg{ProcData} package through a case study of the climate control item in PISA 2012. The dataset \code{cc\_data} is included in the package. In the case study, we explore how well the response outcome can be predicted from the response processes. In particular, we hope to understand what behavior patterns in response processes are closely related to answering the item correctly. 

\subsection{Preparation}
We randomly choose a subset of size 3000 from dataset \code{cc\_data} for our analysis.
\begin{Schunk}
\begin{Sinput}
R> set.seed(12345)
R> n <- 3000
R> idx <- sample(1:length(cc_data$responses), n)
R> seqs <- sub_seqs(cc_data$seqs, idx)
R> y <- cc_data$responses[idx]
\end{Sinput}
\end{Schunk}
In the subset, about 53\% of the respondents answered the item correctly.
\begin{Schunk}
\begin{Sinput}
R> table(y) / n # proportion of incorrect (0) and correct (1) answers
\end{Sinput}
\begin{Soutput}
y
        0         1 
0.4656667 0.5343333 
\end{Soutput}
\end{Schunk}
We take a glance at the response processes by summarizing them through the \code{summary} method for \class{proc} objects.
\begin{Schunk}
\begin{Sinput}
R> seqs_summary <- summary(seqs)
R> seqs_summary$n_action # number of unique actions
\end{Sinput}
\begin{Soutput}
[1] 128
\end{Soutput}
\begin{Sinput}
R> seqs_summary$actions # action set
\end{Sinput}
\begin{Soutput}
  [1] "-1_-1_-1" "-1_-1_-2" "-1_-1_0"  "-1_-1_1"  "-1_-1_2" 
  [6] "-1_-2_-1" "-1_-2_-2" "-1_-2_0"  "-1_-2_1"  "-1_-2_2" 
 [11] "-1_0_-1"  "-1_0_-2"  "-1_0_0"   "-1_0_1"   "-1_0_2"  
 [16] "-1_1_-1"  "-1_1_-2"  "-1_1_0"   "-1_1_1"   "-1_1_2"  
 [21] "-1_2_-1"  "-1_2_-2"  "-1_2_0"   "-1_2_1"   "-1_2_2"  
 [26] "-2_-1_-1" "-2_-1_-2" "-2_-1_0"  "-2_-1_1"  "-2_-1_2" 
 [31] "-2_-2_-1" "-2_-2_-2" "-2_-2_0"  "-2_-2_1"  "-2_-2_2" 
 [36] "-2_0_-1"  "-2_0_-2"  "-2_0_0"   "-2_0_1"   "-2_0_2"  
 [41] "-2_1_-1"  "-2_1_-2"  "-2_1_0"   "-2_1_1"   "-2_1_2"  
 [46] "-2_2_-1"  "-2_2_-2"  "-2_2_0"   "-2_2_1"   "-2_2_2"  
 [51] "0_-1_-1"  "0_-1_-2"  "0_-1_0"   "0_-1_1"   "0_-1_2"  
 [56] "0_-2_-1"  "0_-2_-2"  "0_-2_0"   "0_-2_1"   "0_-2_2"  
 [61] "0_0_-1"   "0_0_-2"   "0_0_0"    "0_0_1"    "0_0_2"   
 [66] "0_1_-1"   "0_1_-2"   "0_1_0"    "0_1_1"    "0_1_2"   
 [71] "0_2_-1"   "0_2_-2"   "0_2_0"    "0_2_1"    "0_2_2"   
 [76] "1_-1_-1"  "1_-1_-2"  "1_-1_0"   "1_-1_1"   "1_-1_2"  
 [81] "1_-2_-1"  "1_-2_-2"  "1_-2_0"   "1_-2_1"   "1_-2_2"  
 [86] "1_0_-1"   "1_0_-2"   "1_0_0"    "1_0_1"    "1_0_2"   
 [91] "1_1_-1"   "1_1_-2"   "1_1_0"    "1_1_1"    "1_1_2"   
 [96] "1_2_-1"   "1_2_-2"   "1_2_0"    "1_2_1"    "1_2_2"   
[101] "2_-1_-1"  "2_-1_-2"  "2_-1_0"   "2_-1_1"   "2_-1_2"  
[106] "2_-2_-1"  "2_-2_-2"  "2_-2_0"   "2_-2_1"   "2_-2_2"  
[111] "2_0_-1"   "2_0_-2"   "2_0_0"    "2_0_1"    "2_0_2"   
[116] "2_1_-1"   "2_1_-2"   "2_1_0"    "2_1_1"    "2_1_2"   
[121] "2_2_-1"   "2_2_-2"   "2_2_0"    "2_2_1"    "2_2_2"   
[126] "end"      "reset"    "start"   
\end{Soutput}
\begin{Sinput}
R> range(seqs_summary$seq_length) # range of sequence lengths
\end{Sinput}
\begin{Soutput}
[1]   3 183
\end{Soutput}
\begin{Sinput}
R> # action transition probability matrix
R> trans_mat <- seqs_summary$trans_count
R> trans_mat <- trans_mat / rowSums(trans_mat) 
\end{Sinput}
\end{Schunk}
\begin{figure}[t!]
\centering
\includegraphics{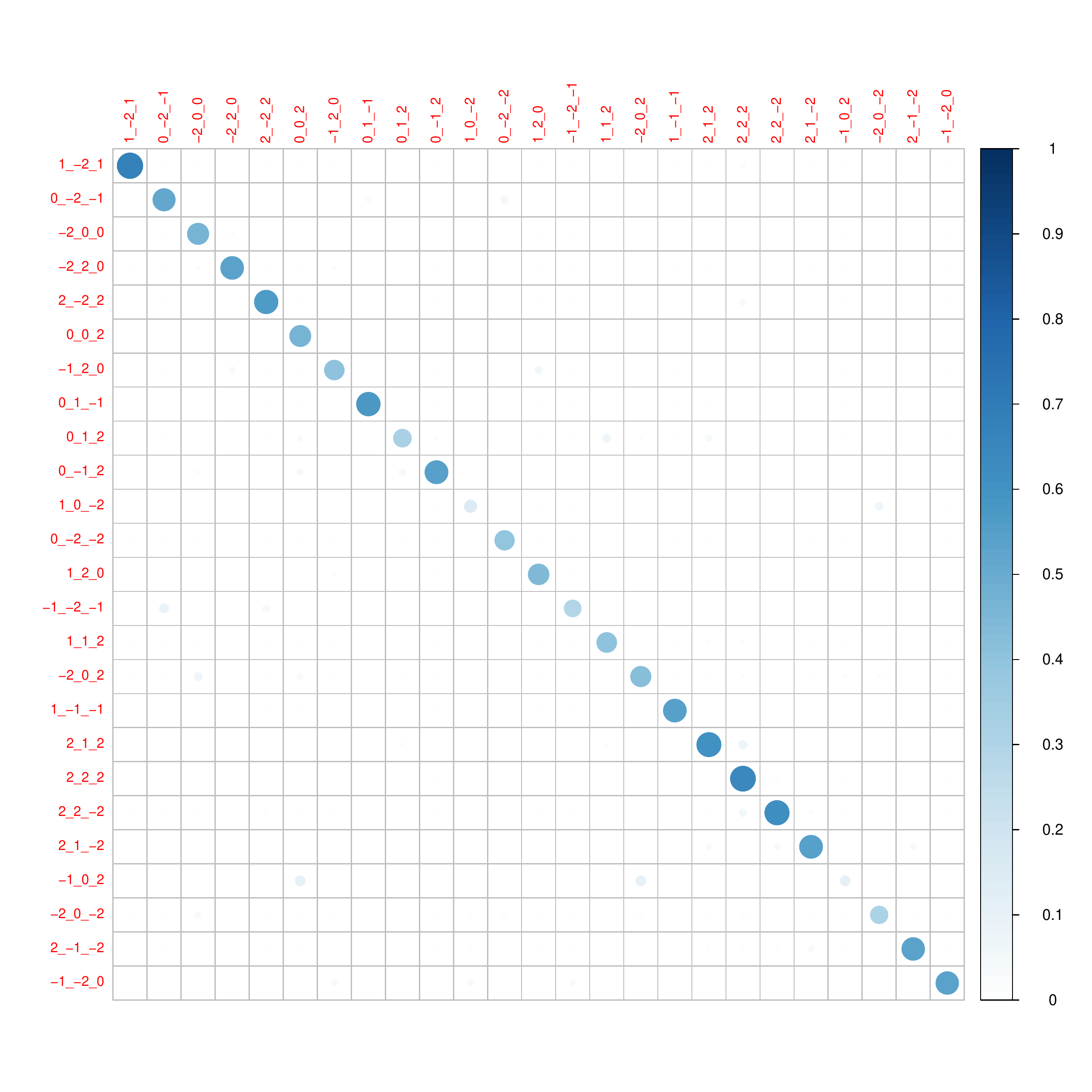}
\caption{\label{fig:trans_mat} Action transition probability matrix.}
\end{figure}
Figure \ref{fig:trans_mat} present a $25 \times 25$ submatrix of the $128\times 128$ action transition probability matrix \code{trans_mat}. The 25 actions corresponds to the submatrix are randomly selected from the action set. The dark diagonal of the submatrix indicates that respondents tend to repeat an action several times before taking a different action. To avoid the repeated actions diluting the information in other actions and to make the response processes less noisy, we remove the repeated actions by using function \code{remove\_repeat}. 
\begin{Schunk}
\begin{Sinput}
R> seqs <- remove_repeat(seqs)
\end{Sinput}
\end{Schunk}

\subsection{Outcome Prediction Based on Extracted Features}
As we mentioned earlier, response processes are in a nonstandard format and thus cannot be easily incorporated in traditional statistical models such as linear models and generalized linear models. To make use of these classical tools for studying the relation between response processes and response outcomes, we first compress the information in the response processes into features, which are in a standard matrix format, and then use them as covariates to fit a logistic regression.

\begin{figure}[t!]
\includegraphics{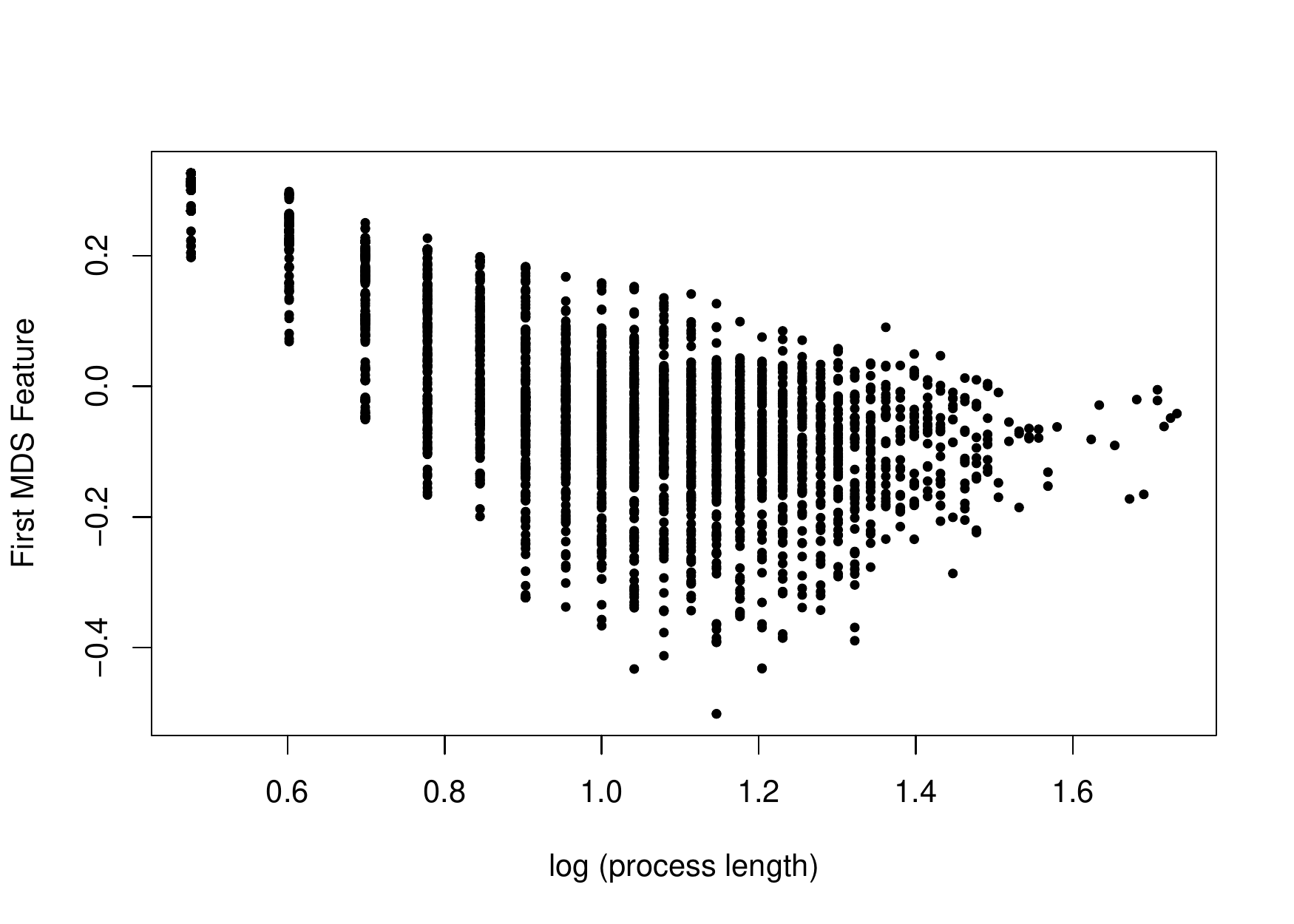}
\caption{\label{fig:feature1} First mds feature against the logarithm of process length.}
\end{figure}

We start by extracting features through multidimensional scaling. The number of features to be extracted is chosen by five-fold cross-validation. The candidate values of the number of features are $10, 20, \ldots, 100$. We only consider the action sequences here and set \code{dist_type = "oss_action"}. Note that by setting \code{return_dist = TRUE}, we ask \code{chooseK_mds} to return the dissimilarity matrix for future use to avoid repeated calculations. The returned dissimilarity matrix and the selected number of features are then passed to \code{seq2feature_mds} for feature extraction.
\begin{Schunk}
\begin{Sinput}
R> mds_K_res <- chooseK_mds(seqs, 1:10*10, dist_type = "oss_action", 
+                           return_dist = TRUE)
R> K <- mds_K_res$K
R> K
\end{Sinput}
\begin{Soutput}
[1] 60
\end{Soutput}
\begin{Sinput}
R> mds_res <- seq2feature_mds(mds_K_res$dist_mat, K = K)
\end{Sinput}
\end{Schunk}
The 60 extracted features can then be used as covariates in logistic regression. To evaluate the out-of-sample prediction performance, the dataset is split into training, validation, and test sets.
\begin{Schunk}
\begin{Sinput}
R> n_train <- 2000
R> n_valid <- 500
R> n_test <- 500
R> index_train <- sample(1:n, n_train)
R> index_valid <- sample(setdiff(1:n, index_train), n_valid)
R> index_test <- sample(1:n, c(index_train, index_valid))
\end{Sinput}
\end{Schunk}
The fitting and prediction of logistic regression models can be achieved by the \code{glm} function and the \code{predict} method as usual. 
\begin{Schunk}
\begin{Sinput}
R> mds_data <- data.frame(y = y, x = mds_res$theta)
R> mds_glm_res <- glm(y ~ ., family = "binomial", 
+                     subset=c(index_train, index_valid), data=mds_data)
R> yhat_test <- as.numeric(predict(mds_glm_res, newdata = mds_data[index_test,]) > 0)
R> # out-of-sample prediction accuracy
R> mean(y[index_test] == yhat_test)
\end{Sinput}
\begin{Soutput}
[1] 0.7981735
\end{Soutput}
\begin{Sinput}
R> summary(mds_glm_res)
\end{Sinput}
\begin{Soutput}
Call:
glm(formula = y ~ ., family = "binomial", data = mds_data, subset = c(index_train, 
    index_valid))

Deviance Residuals: 
    Min       1Q   Median       3Q      Max  
-3.7806  -0.6561   0.1878   0.7678   2.2290  

Coefficients:
            Estimate Std. Error z value Pr(>|z|)    
(Intercept)  0.23119    0.05255   4.399 1.09e-05 ***
x.1         -9.87064    0.41848 -23.587  < 2e-16 ***
x.2         -0.73587    0.37248  -1.976 0.048201 *  
x.3         -1.68177    0.48546  -3.464 0.000532 ***
x.4          5.42484    0.49103  11.048  < 2e-16 ***
x.5          1.38222    0.51211   2.699 0.006954 ** 
x.6         -1.10816    0.63259  -1.752 0.079814 .  
x.7          1.25793    0.66810   1.883 0.059721 .  
x.8         -3.68047    0.68458  -5.376 7.61e-08 ***
x.9          1.64550    0.82766   1.988 0.046796 *  
x.10         2.18376    0.83807   2.606 0.009169 ** 

(output omitted)
   
---
Signif. codes:  0 '***' 0.001 '**' 0.01 '*' 0.05 '.' 0.1 ' ' 1

(Dispersion parameter for binomial family taken to be 1)

    Null deviance: 3457.4  on 2499  degrees of freedom
Residual deviance: 2382.7  on 2439  degrees of freedom
AIC: 2504.7

Number of Fisher Scoring iterations: 5
\end{Soutput}
\end{Schunk}
Since the out-of-sample prediction accuracy (0.80) is much higher than the benchmark (0.54), which is the accuracy if we naively predict that all respondents in the test set answered the item correctly, it is reasonable to believe that some behavior patterns in the response processes are closely related to the response outcomes. 
By inspecting the \code{glm} output and refiting a logistic regression with the first feature as the only covariate, we find that including only the first feature can already produce a prediction accuracy of 0.77. 
\begin{Schunk}
\begin{Sinput}
R> mds_glm_res1 <- glm(y ~ x.1, family = "binomial", 
+                      subset=c(index_train, index_valid), data=mds_data)
R> yhat_test1 <- as.numeric(predict(mds_glm_res1, newdata = mds_data[index_test,]) > 0)
R> mean(y[index_test] == yhat_test1)
\end{Sinput}
\begin{Soutput}
[1] 0.7730594
\end{Soutput}
\end{Schunk}
Now we examine the behavior patterns associated with the first multidimensional scaling feature. We order the response processes according to the value of their first feature and use the \code{print} method for class \class{proc} to display the response processes with the highest and lowest values.
\begin{Schunk}
\begin{Sinput}
R> o_mds1 <- order(mds_res$theta[,1])
R> # response processes corresponding to the highest feature values
R> print(seqs, index = head(o_mds1))
\end{Sinput}
\begin{Soutput}
'proc' object of  3000  processes


FRA000019103869 
      Step 1 Step 2 Step 3 Step 4 Step 5 Step 6 Step 7 Step 8 Step 9
Event start  1_0_0  2_0_0  -1_0_0 -2_0_0 0_1_0  0_2_0  0_-1_0 0_-2_0
Time   0.0   36.6   40.9   46.6   50.3   59.4   64.0   66.3   69.4  
      Step 10 Step 11 Step 12 Step 13 Step 14
Event 0_0_1   0_0_2   0_0_-1  0_0_-2  end    
Time  77.7    80.3    83.9    86.0    93.2   

AUT000007001735 
      Step 1 Step 2 Step 3 Step 4 Step 5 Step 6 Step 7 Step 8 Step 9
Event start  1_0_0  2_0_0  -2_0_0 -1_0_0 0_1_0  0_2_0  0_-1_0 0_-2_0
Time    0.0   57.4   60.6   62.8   67.1   79.6   81.1   83.3   85.3 
      Step 10 Step 11 Step 12 Step 13 Step 14
Event 0_0_1   0_0_2   0_0_-1  0_0_-2  end    
Time   97.2    98.6   100.7   102.5   119.5  

PRT000009602769 
      Step 1 Step 2 Step 3 Step 4 Step 5 Step 6 Step 7 Step 8 Step 9
Event start  1_0_0  2_0_0  -1_0_0 -2_0_0 0_1_0  0_2_0  0_-1_0 0_0_1 
Time    0.0   73.1   82.5   89.2   94.3  109.5  114.8  119.8  128.8 
      Step 10 Step 11
Event 0_0_2   end    
Time  131.4   139.7  

BRA000083118946 
      Step 1 Step 2 Step 3 Step 4 Step 5 Step 6 Step 7 Step 8 Step 9
Event start  1_0_0  2_0_0  -1_0_0 -2_0_0 reset  0_1_0  0_2_0  0_-1_0
Time    0.0  141.7  151.2  157.7  164.6  182.1  185.1  188.8  194.2 
      Step 10 Step 11 Step 12 Step 13 Step 14 Step 15 Step 16
Event 0_-2_0  reset   0_0_1   0_0_2   0_0_-1  0_0_-2  end    
Time  196.3   204.0   206.6   209.4   215.9   218.6   240.4  

KOR000006202016 
      Step 1 Step 2 Step 3 Step 4 Step 5 Step 6 Step 7 Step 8 Step 9
Event start  2_0_0  1_0_0  -1_0_0 -2_0_0 reset  0_1_0  0_2_0  0_-1_0
Time   0.0   46.1   52.4   56.4   59.2   62.5   67.3   69.3   70.3  
      Step 10 Step 11 Step 12 Step 13 Step 14 Step 15 Step 16
Event 0_-2_0  reset   0_0_1   0_0_2   0_0_-1  0_0_-2  end    
Time  72.4    78.9    81.1    84.1    86.6    89.0    94.6   
\end{Soutput}
\begin{Sinput}
R> # response processes corresponding to the lowest feature values
R> print(seqs, index = tail(o_mds1))
\end{Sinput}
\begin{Soutput}
'proc' object of  3000  processes


KOR000010803463 
      Step 1 Step 2 Step 3
Event start  0_0_0  end   
Time   0.0   50.6   63.9  

MYS000009803099 
      Step 1 Step 2 Step 3
Event start  0_0_0  end   
Time    0.0   38.4  114.8 

URY000004601383 
      Step 1 Step 2 Step 3
Event start  0_0_0  end   
Time   0.0   61.0   66.9  

MYS000001800562 
      Step 1 Step 2 Step 3
Event start  0_0_0  end   
Time   0.0   70.4   84.5  

MYS000015604929 
      Step 1 Step 2 Step 3
Event start  0_0_0  end   
Time    0.0   71.5  107.1 
\end{Soutput}
\end{Schunk}
The response processes corresponding to the highest feature values are often very short, meaning that little meaningful interaction with computer interface is made. Respondents with this type of behaviors are unlikely to answer the question correctly. The response processes corresponding to the lowest feature values are often longer. However, their lengths are not the longest according to Figure \ref{fig:feature1} where the value of the first feature is plotted against the logarithm of the length of the corresponding process. A closer look at the response processes with the lowest values of the feature reveals that these respondents often explore the function of one control bar at a time. This ``varying-one-thing-at-a-time'' strategy is efficient to get the correct answer.

We can also extract features by constructing a sequence autoencoder. 
Below is an example of extracting action sequence autoencoder features by \code{seq2feature_seq2seq} in \pkg{ProcData}. 
\begin{Schunk}
\begin{Sinput}
R> ae_a_res <- seq2feature_seq2seq(seqs, ae_type = "action", K = K, 
+                                rnn_type = "gru", samples_train = index_train, 
+                                samples_valid = index_valid, verbose = FALSE)
\end{Sinput}
\end{Schunk}
The timestamp sequence can easily be taken into account by setting \code{ae_type = "both"}.
\begin{Schunk}
\begin{Sinput}
R> ae_at_res <- seq2feature_seq2seq(seqs, ae_type = "both", K = K, 
+                                rnn_type = "gru", samples_train = index_train, 
+                                samples_valid = index_valid, verbose = FALSE)
\end{Sinput}
\end{Schunk}
The logarithm of the inter-arrival time sequence is used here as by default \code{cumulative = FALSE} and \code{log = TRUE}.
Similar to the multidimensional scaling features, features extracted by sequence autoencoders can be used as covariates in classical models. In our case, we can fit logistic regression models for predicting response outcomes.
\begin{Schunk}
\begin{Sinput}
R> ae_a_data <- data.frame(y = y, x = ae_a_res$theta)
R> ae_glm_a_res <- glm(y ~ ., family = "binomial", 
+                      subset=c(index_train, index_valid), data=ae_a_data)
R> yhat_test_a <- as.numeric(predict(ae_glm_a_res, 
+                                    newdata = ae_a_data[index_test,]) > 0)
R> # out-of-sample prediction accuracy (action only)
R> mean(y[index_test] == yhat_test_a)
\end{Sinput}
\begin{Soutput}
[1] 0.8401826
\end{Soutput}
\begin{Sinput}
R> ae_at_data <- data.frame(y = y, x = ae_at_res$theta)
R> ae_glm_at_res <- glm(y ~ ., family = "binomial", 
+                       subset=c(index_train, index_valid), data=ae_at_data)
R> yhat_test_at <- as.numeric(predict(ae_glm_at_res, 
+                                     newdata = ae_at_data[index_test,]) > 0)
R> # out-of-sample prediction accuracy
R> mean(y[index_test] == yhat_test_at)
\end{Sinput}
\begin{Soutput}
[1] 0.8424658
\end{Soutput}
\end{Schunk}
The prediction accuracy is about 5\% higher than that obtained by MDS features. Important patterns in response processes can be examined similarly. We refer the interested readers to \citet{tang2019seq2seq}.

\subsection{Feature Extraction with Targeted Variable}
If prediction is the only goal for exploring the relationship between response processes and a binary or numeric response variable, one can achieve this by fitting a sequence model. We still use the response outcome as the response variable of our interest to illustrate how this can be done in \pkg{ProcData}.

\begin{figure}[t!]
\centering
\includegraphics{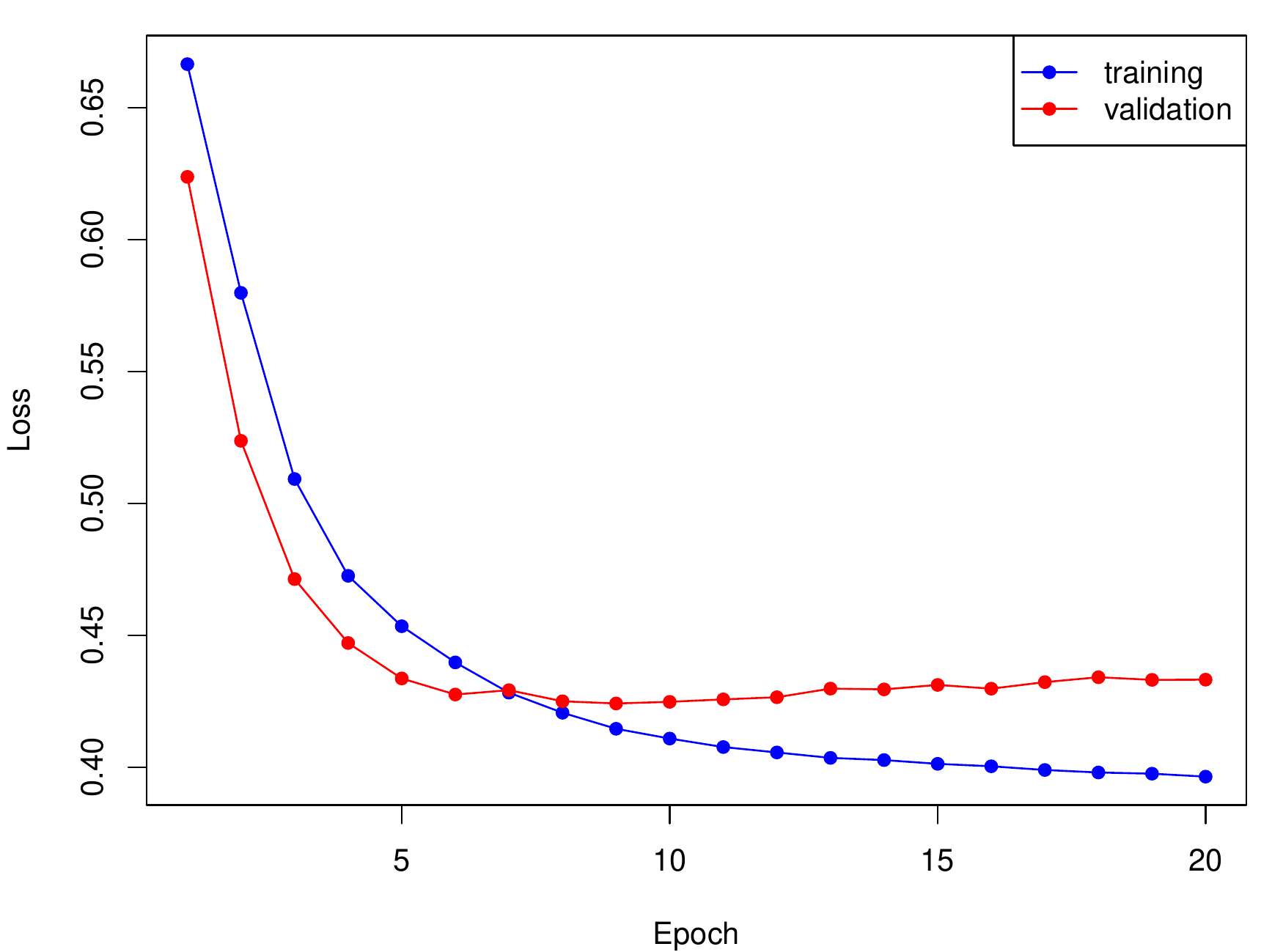}
\caption{\label{fig:seqm_trace} Trace of training and validation loss.}
\end{figure}

We split the data into training, validation, and testing set as before. The training and validation sets are used for fitting the sequence model while the testing set is used for evaluating its prediction performance. The response processes used for model fitting and testing are obtained by subsetting the \class{proc} object \code{seqs} with function \fct{sub\_seqs} in \pkg{ProcData}.
\begin{Schunk}
\begin{Sinput}
R> seqs_train <- sub_seqs(seqs, c(index_train, index_valid))
R> seqs_test <- sub_seqs(seqs, index_test)
R> y_train <- y[c(index_train, index_valid)]
R> y_test <- y[index_test]
\end{Sinput}
\end{Schunk}
We first consider the task of predicting the response outcome from the action sequence in a response process. The sequence model that fulfills this task can be fitted by calling function \fct{seqm}. Since the response outcome is a binary variable, we specify \code{response_type = "binary"}. \code{n_epoch} is the total number of epochs to be run for estimating the parameters. The fitted model is the one that produces the lowest loss on the validation set, which is specified by passing the indices of the processes through \code{index_valid}.
\begin{Schunk}
\begin{Sinput}
R> seqm_res <- seqm(seqs_train, y_train, response_type = "binary",
+                   K_emb = 5, K_rnn = 5, n_epoch = 20, 
+                   max_len = max(seqs_summary$seq_length), 
+                   index_valid = n_train + 1:n_valid)
\end{Sinput}
\end{Schunk}
The convergence of training process can be examined by a plot of the value of the loss function at the end of each epoch stored in \code{seqm_res$history} (Figure \ref{fig:seqm_trace}). 
Prediction based on the fitted model can be made by calling the \code{predict} method for \class{seqm} objects.
\begin{Schunk}
\begin{Sinput}
R> seq_pred_res <- predict(seqm_res, new_seqs = seqs_test)
R> mean(as.numeric(seq_pred_res > 0.5) == y_test)
\end{Sinput}
\begin{Soutput}
[1] 0.847032
\end{Soutput}
\end{Schunk}
The prediction accuracy is slightly higher than that obtained by logistic regression on extracted features.

A sequence model incorporating the information in the timestamp sequences can be fitted by setting \code{include_time = TRUE} in \fct{seqm}. Prediction from the fitted model can be made in the same way as before.
\begin{Schunk}
\begin{Sinput}
R> seqm_res2 <- seqm(seqs_train, y_train, response_type = "binary", 
+                    include_time = TRUE, K_emb = 5, K_rnn = 5, n_epoch = 20, 
+                    max_len = max(seqs_summary$seq_length),
+                    index_valid = n_train + 1:n_valid)
R> seq_pred_res2 <- predict(seqm_res2, new_seqs = seqs_test)
R> mean(as.numeric(seq_pred_res2 > 0.5) == y_test)
\end{Sinput}
\begin{Soutput}
[1] 0.656621
\end{Soutput}
\end{Schunk}
Adding timestamp sequences impairs the prediction accuracy, indicating that timestamp sequences of the climate control item do not contain much additional information about the response outcomes.

In many cases, both the response processes and background information such as age, education level, and employment status are considered in prediction. 
These background information can be added to the sequence model as covariates supplied through \code{covariates} in \code{seqm}. Covariates for the test set should be supplied to \fct{predict.seq} through \code{new_covariates} when we make predictions. Since background information is not available in the climate control dataset \code{cc_data}, we demonstrate it by adding the first five MDS features.
\begin{Schunk}
\begin{Sinput}
R> X <- mds_res$theta[,1:5]
R> X_train <- X[c(index_train, index_valid), ]
R> X_test <- X[index_test, ]
R> seqm_res3 <- seqm(seqs_train, y_train, covariates = X_train, 
+                    response_type = "binary", K_emb = 5, K_rnn = 5, 
+                    n_epoch = 20, max_len = max(seqs_summary$seq_length),
+                    index_valid = n_train + 1:n_valid)
R> seq_pred_res3 <- predict(seqm_res3, new_seqs = seqs_test, 
+                           new_covariates = X_test)
R> mean(as.numeric(seq_pred_res3 > 0.5) == y_test)
\end{Sinput}
\begin{Soutput}
[1] 0.847032
\end{Soutput}
\end{Schunk}

Prediction accuracy does not change significantly after the MDS features are incorporated. This indicates that the action sequence information contained in the features are well-captured in the sequence model.


\section{Summary}\label{sec:summary}
To the authors' best knowledge, \pkg{ProcData} is the first package designed for process data analysis. It can be used for analyzing process data generating from a wide range of human-computer interface. \pkg{ProcData} includes an \proglang{S}3 class for response processes and functions for processing, describing, and analyzing process data. Two feature extraction methods and fitting and making predition from neural-network-based sequence models are implemented in \pkg{ProcData}. These tools are easy to use. Users do not need to handle the construction and training of neural network if they want to use neural network related models.
Process data analysis is an active and quickly rising field. We envision to include more state-of-the-art methods for analyzing response processes in future versions of \pkg{ProcData}. 

\section*{Acknowledgments}

This research is supported in part by NSF IIS-1633360 and NSF SES-1826540.


\bibliography{refs}

%
%
%
%
%
%
%


\end{document}